\documentclass[11pt,a4paper]{article}
\usepackage{jheppub}

\usepackage{amsmath,amssymb,esint}
\usepackage[normalem]{ulem}
\usepackage{graphicx}
\usepackage{float}
\usepackage{bbold}
\usepackage{enumitem}

\usepackage{amssymb}
\usepackage{amsthm}
\theoremstyle{plain}
\numberwithin{equation}{section}
\usepackage{multirow}
\usepackage{wasysym}
\usepackage{dynkin-diagrams}
\usepackage{arydshln}
\usepackage{mathtools}
\DeclarePairedDelimiter{\ceil}{\lceil}{\rceil}

\usepackage{esvect}
\usepackage{physics}
\usepackage{tensor}
\usepackage{bm}
\usepackage{color}
\usepackage{hyperref}
\usepackage{slashed}
\usepackage{blkarray}

\newcommand{\cG}{\mathcal{G}}
\newcommand{\uv}{\mathbf{e}}
\newcommand{\bn}{\mathbf{n}}
\newcommand{\beps}{\boldsymbol{\epsilon}}
\newcommand{\bq}{\mathbf{q}}
\newcommand{\bk}{\mathbf{k}}
\newcommand{\cM}{\mathcal{M}}
\newcommand{\nf}{\mathfrak{f}}

\usepackage{listings}

\usepackage{tikz}
\usetikzlibrary{arrows,matrix,positioning,shapes,arrows}
\usetikzlibrary{shapes.geometric, arrows, calc, intersections}
\usetikzlibrary{positioning}

\usepackage{youngtab}
\usepackage{changepage}

\newtheorem{lem}{Lemma}[section]

\newtheorem{cor}{Corollary}[section]

\newcommand*\xbar[1]{%
  \hbox{%
    \vbox{%
      \hrule height 0.5pt % The actual bar
      \kern0.4ex%         % Distance between bar and symbol
      \hbox{%
        \kern-0.25em%      % Shortening on the left side
        \ensuremath{#1}%
        \kern-0.1em%      % Shortening on the right side
      }%
    }%
  }%
}

\title{Quark Masses and Mixing in String-Inspired Models}
\author[a]{Andrei Constantin,}
\author[a]{Cristofero S.\ Fraser-Taliente,}
\author[b,c]{Thomas R.\ Harvey,}
\author[a]{Lucas T. Y.\ Leung,}
\author[a]{Andre Lukas}

\affiliation[a]{Rudolf Peierls Centre for Theoretical Physics, University of Oxford, Parks Road, Oxford OX1 3PU, UK}
\affiliation[b]{The NSF AI Institute for Artificial Intelligence and Fundamental Interactions,  Massachusetts Institute of Technology, 77 Massachusetts Avenue,  Cambridge, MA 02139, USA}

\affiliation[c]{Center for Theoretical Physics, Massachusetts Institute of Technology, 77 Massachusetts Avenue,  Cambridge, MA 02139, USA}

\emailAdd{andrei.constantin@physics.ox.ac.uk}
\emailAdd{cristofero.fraser-taliente@physics.ox.ac.uk}
\emailAdd{trharvey@mit.edu}
\emailAdd{lucas.leung@physics.ox.ac.uk}
\emailAdd{andre.lukas@physics.ox.ac.uk}

\abstract{We study a class of supersymmetric Froggatt-Nielsen (FN) models with multiple $U(1)$ symmetries and Standard Model (SM) singlets inspired by heterotic string compactifications on Calabi-Yau threefolds. The string-theoretic origin imposes a particular charge pattern on the SM fields and FN singlets, dividing the latter into perturbative and non-perturbative types. Employing systematic and heuristic search strategies, such as genetic algorithms, we identify charge assignments and singlet VEVs that replicate the observed mass and mixing hierarchies in the quark sector, and subsequently refine the Yukawa matrix coefficients to accurately match the observed values for the Higgs VEV, the quark and charged lepton masses and the CKM matrix.  
This bottom-up approach complements top-down string constructions and our results demonstrate that string FN models possess a sufficiently rich structure to account for flavour physics. On the other hand, the limited number of distinct viable charge patterns identified here indicates that flavour physics imposes tight constraints on string theory models, adding new constraints on particle spectra that are essential for achieving a realistic phenomenology.}

\begin{document}
\date{Last updated: \today}

\maketitle

\newpage
\section{Introduction} \label{sec:intro}

Over the past four decades, the development of string theory models of particle physics has gradually advanced, leading to the successful construction of models with the correct (supersymmetric) particle content of the Standard Model. Various approaches have contributed to this progress, and many such models are now known. For comprehensive reviews and detailed discussions, see~\cite{ibanez_uranga_2012,Marchesano:2024gul,Cvetic:2022fnv} and the references therein.
The next crucial step toward achieving a fully realistic string-derived Standard Model is to obtain the correct masses and mixing of quarks and leptons. Recently, significant progress has been made in this direction within top-down string constructions~\cite{Constantin:2024yxh}, specifically in the context of $E_8\times E_8$ heterotic string compactifications on smooth Calabi-Yau threefolds with abelian internal gauge fluxes\footnote{See also Refs.~\cite{Butbaia:2024tje, Berglund:2024uqv} for similar calculations in the context of standard embedding models.}. 
The calculations presented in Ref.~\cite{Constantin:2024yxh} relied extensively on neural network techniques and represented a substantial undertaking in numerical differential geometry.
A~limited exploration of the complex structure moduli space was also included.\\[2mm]
In this paper, we study fermion masses and mixing from a complementary, bottom-up perspective in the context of string-inspired four-dimensional theories with $\mathcal N=1$ supersymmetry. 
In string models, matter fields are tied to branes, singularities, and internal gauge flux within the underlying manifold. Mathematically, gauge flux is represented by vector bundles over the internal manifold or its submanifolds. Often, at certain loci in their moduli space, these vector bundles can `split', meaning they exhibit a product structure group. At these split loci, Green-Schwarz anomalous $U(1)$ symmetries emerge, which manifest in the low-energy effective field theory as global $U(1)$ symmetries. Moving away from the split loci corresponds to switching on bundle moduli VEVs; in this process, the effects of the $U(1)$ symmetries persist in the moduli space, constraining the allowed operators in the effective field theory~\cite{Buchbinder:2014sya}. 
Quarks and leptons, along with various geometric moduli (bundle moduli and K\"ahler moduli), acquire charges under these symmetries, imposing constraints on the allowed operators in the effective theory.
In this paper, we analyse the implications of these symmetries for fermion masses and mixing. The specific form of the four-dimensional supergravity theories considered here is derived from heterotic Calabi-Yau compactifications with split vector bundles~\cite{Distler:1987ee, Blumenhagen:2006ux, Blumenhagen:2006wj, Anderson:2012yf, Anderson:2011ns, Anderson:2013xka}. The calculation of physical Yukawa couplings in Ref.~\cite{Constantin:2024yxh} has been performed in the same setting and one of the reasons behind the present approach is to guide such calculations towards phenomenologically viable regions of the moduli space, including the bundle moduli space. Furthermore, it is known that related effective theories arise in F-theory compactifications~\cite{Font:2008id, Dudas:2009hu, Krippendorf:2015kta}, where similar methods can also be applied.\\[2mm]
From a four-dimensional model-building perspective, our approach is similar to the Froggatt-Nielsen (FN) mechanism for the origin of the mass hierarchy~\cite{FROGGATT1979277, Leurer:1992wg, Leurer:1993gy, Davidson:1979wr,Davidson:1981zd,Davidson:1983fy,Dudas:1995yu,Dudas:1996fe,Ibanez:1994ig}. However, it incorporates several important differences stemming from the presumed stringy origin of our models. For standard FN models of the simplest type, one assumes the presence of a $U(1)$ flavour symmetry in addition to the Standard Model (SM) gauge group, typically referred to as horizontal $U(1)$ symmetry. Both the SM fields and an additional SM singlet scalar field (flavon or FN scalar) $\phi$ carry charge under the $U(1)$ flavour symmetry. The presence of $U(1)$-invariant operators of the form $\phi^p\times(\text{Yukawa interaction term})$, where $p$ is a non-negative integer, implies that, once $\phi$ develops a vacuum expectation value (VEV) $\langle\phi\rangle$, a Yukawa coupling constant of order $\langle\phi\rangle^p$ is generated. The hope is then that the hierarchy of fermion masses can be explained by a somewhat small value of $\langle\phi\rangle$ (measured in units of an appropriate mass scale) and different integers $p$ for the various entries of the Yukawa matrices, as dictated by the $U(1)$ charges.\\[2mm]
Our string-inspired FN models (or simply ``string FN models'') deviate from the standard FN mechanism in several ways, being more general in some aspects, whilst more specific in others. The string FN models considered in this paper are characterised by:
\begin{description}%[topsep=0pt, partopsep=0pt]
\setlength{\itemsep}{0pt}%
\setlength{\parskip}{0pt}
\item[Presence of supersymmetry.] We assume that our models are MSSM-like with Standard Model gauge group. In addition, the structure of the underlying string models implies that the horizontal $U(1)$ charges of the MSSM multiplets can be organised in terms of $SU(5)$ multiplets. The form of the Yukawa couplings is dictated by supersymmetry, the SM gauge symmetry and the horizontal symmetry. 
\item[Underlying GUT symmetry.]
While our models have a standard model gauge group and are by no means field theory GUTs there is an underlying $SU(5)$ GUT symmetry at the string level which leaves certain traces in the low-energy theory. In particular, the FN $U(1)$ symmetries commute with the $SU(5)$ GUT symmetry so that $U(1)$ charges can be organised by GUT multiplets. 
\item[Several horizontal $U(1)$s.] We consider symmetry groups of the form $\cG\cong U(1)^{f-1}$, where $f=2,3,4$ or $5$.  It is convenient to parametrize $\cG$ by starting with a partition $\mathbf{n} = \left(n_1,...,n_{f}\right)$ of $5$ which, in the underlying string model, describes the splitting type of rank $5$ vector bundles. In terms of this partition, the group $\cG$ is given by
\begin{equation} \label{Gdef_intro}
	\cG = \left\{ \left(e^{i\epsilon^1},...,e^{i\epsilon^f}\right) \,\middle\vert\, \epsilon^a\in\mathbb{R}\,,\; \bn\cdot\beps=0  \right\}\cong U(1)^{f-1}\; ,
\end{equation}
where $\beps=(\epsilon^1,\ldots,\epsilon^f)$ are the group parameters. Representations of $\cG$ are labelled by integer vectors $\bq=(q_1,\ldots ,q_f)\in\mathbb{Z}^f$, subject to the identification 
\begin{equation}\label{equiv_intro}
\bq\sim \bq'\quad\Leftrightarrow\quad
\bq-\bq'\in\mathbb{Z}\bn\; .
\end{equation}
\item[Several FN scalars.] 
Instead of a single FN scalar, we introduce multiple $\cG$-charged scalars (chiral superfields) which come in two varieties. We denote the first type of FN scalars (chiral SM singlets) as $\phi^{(\alpha)}$, where $\alpha$ is now an index, not a power. In a string theory context, these are interpreted as bundle moduli and contribute to the superpotential and the K\"ahler potential at perturbative level. Denoting by $\mathbf{q}(\phi^{(\alpha)})$ the corresponding $\cG$-charge vector, the singlet  $\phi^{(\alpha)}$ transforms linearly as
\begin{equation}
\phi^{(\alpha)}\longmapsto e^{-i \mathbf{q}(\phi^{(\alpha)})\cdot \beps }\phi^{(\alpha)}~,
\end{equation}
where the vector $\beps$ contains the $U(1)$ group parameters. Additionally, we introduce FN scalars $\Phi^{(i)}=e^{-T^i}$ which feature in non-perturbative string contributions to the superpotential and the K\"ahler potential.\footnote{These contributions are associated with worldsheet instantons and arise as effects of the string worldsheet wrapping holomorphic curves inside the Calabi-Yau threefold. The instantons contribute exponentially suppressed terms, where the exponent is proportional to the volume of the wrapped curves, as measured by the K\"ahler moduli.} These are associated with geometrical K\"ahler moduli fields $T^i$ which transform non-linearly under $\cG$:
\begin{equation}\label{T_transf}
T^i\mapsto T^i+i\mathbf{k}^i\cdot\beps~,    
\end{equation}
where $\mathbf{k}^i$ are integral vectors dictated, in the string context, by the topological properties of the split bundle. This non-linear transformation for the K\"ahler moduli $T^i$ implies a linear transformation for the FN scalars $\Phi^{(i)}$. Thus, both types of FN scalars are $\cG$-charged and their presence as singlet insertions is constrained.\footnote{One could be tempted to think of the $U(1)$-charges in this context by analogy with the modular weights that arise in orbifold compactifications, however, this is not helpful. The crucial difference is that in orbifold compactifications the modular weights of low-energy fields (including the K\"ahler moduli) are related to the orbifold geometry and its discrete symmetries. The would-be analogue in smooth Calabi-Yau compactifications corresponds to large diffeomorphisms, e.g.~in the form of discrete symmetries. By contrast, the $U(1)$-symmetries used here arise from gauge transformations in the presence of line bundles.}
\end{description}
So far, the discussion has been fairly general. However, our models are special due to the particular structure of the allowed $\cG$ charges dictated by the presumed underlying string setting.
\begin{description}
    \item[Pattern of charges.]  The SM fields and the perturbative FN scalars $\phi^{(\alpha)}$ can carry charges $\pm 1$ under at most two $U(1)$ symmetries (and are uncharged under the others), according to a certain pattern suggested by the underlying string models and  described in detail in the following section.
\item[Anomaly cancellation.] The charges for the non-perturbative FN scalars $\Phi^{(i)}$ are constrained by anomaly cancellation, which imposes a restriction on the allowed integral vectors $\mathbf{k}^i$ in the transformation~\eqref{T_transf}.
\item[Yukawa unification.] In contrast to traditional field theory GUTs, the string GUT symmetry underlying our models do not automatically impose Yukawa unification~\cite{green_2012_superstring2} since the SM multiplets, which are unified in traditional field theory GUTs, originate from different multiplets in the underlying string theory GUT. However, as mentioned above, these multiplets necessarily share the same $U(1)$ charges. This means our models have a crude form of Yukawa unification where allowed singlet insertions between the d-quark and the charged lepton sector are the same but the coefficients multiplying these insertions do not respect unification and will generically break it.
\end{description}

\noindent {\bfseries Example model.} Before outlining our strategy for identifying such string FN models, it is useful to present an example. The model is characterised by a horizontal symmetry $\cG\cong U(1)^4$, corresponding to the uniform partition $\mathbf{n}=(1,1,1,1,1)$ of $5$. As mentioned above, it is convenient to describe $\cG$ in terms of $5$ group parameters with one relation, such that the charge vectors $\bq=(q_1,\ldots ,q_5)\in\mathbb{Z}^5$ are subject to the identification 
\begin{equation}
\bq\sim \bq'\quad\Leftrightarrow\quad
\bq-\bq'\in\mathbb{Z} (1,1,1,1,1) .
\end{equation} The spectrum contains the Standard Model matter fields, written in $SU(5)$ GUT notation:
\begin{equation}\label{SM_spectrum_intro}
    \begin{tabular}{cccccccc}
$\bm{10}_1$ & $\bm{10}_2$ & $\bm{10}_5$ & $\bm{\bar{5}}_{1,2}$ & $\bm{\bar{5}}_{1,2}$ & $\bm{\bar{5}}_{1,2}$ & $H^u_{4,5}$ & $H^d_{4,5}$~,
\end{tabular}
\end{equation}
as well as five types of perturbative FN scalars 
\begin{equation}
    \begin{tabular}{ccccccc}
$\phi_{5,1}$ & $\phi_{3,5}$ & $\phi_{4,5}$ & $\phi_{1,2}$ & $\phi_{4,1}$~,
\end{tabular}
\end{equation}
and no non-perturbative FN scalars. The subscripts indicate the $U(1)$ charges as follows: 
\begin{equation}
    \mathbf{q}(\mathbf{10}_a) = \mathbf{e}_a~,~~~ \mathbf{q}(\mathbf{\bar{5}}_{a,b}) = -\mathbf{q}(H^u_{a,b})=\mathbf{q}({H}^d_{a,b}) =\mathbf{e}_a+\mathbf{e}_b~,~~~ \mathbf{q}(\phi_{a,b}) = \mathbf{e}_a-\mathbf{e}_b~,
\end{equation}
where $\mathbf{e}_a$ is the $a$-th standard basis vector in 5 dimensions. Our task now is to determine the structure of the Yukawa matrices as dictated by the $U(1)$ symmetries.\\[2mm]
The generic form of the Yukawa couplings is the following:
\begin{equation}
    \begin{aligned}
        \text{up sector:}&~~~~(\text{singlet insertions})\times~  H^u_{-\mathbf{e}_a-\mathbf{e}_b} \bm{10}^i_{\mathbf{e}_c}\bm{10}^j_{\mathbf{e}_d}\\
        \text{down sector:}&~~~~(\text{singlet insertions})\times~  H^d_{\mathbf{e}_a+\mathbf{e}_b} \bar{\mathbf{5}}^i_{\mathbf{e}_c+\mathbf{e}_d}\bm{10}^j_{\mathbf{e}_e}~,
    \end{aligned}
\end{equation}
The leading-order singlet insertions that lead to $\cG$-invariant operators are:
\begin{equation} \label{eq:yuk_eg_1}
\begin{aligned}
\text{up sector:}&~~~~ \begin{pmatrix}
\!\!\!\phi_{5,1}\phi_{4,1}& \phi_{5,1}\phi_{1,2}\phi_{4,1} & ~~\phi_{4,1}\\
\!\!\!~~\phi_{5,1}\phi_{1,2}\phi_{4,1}~~ & ~~\phi_{5,1}\phi_{1,2}^2\phi_{4,1}~~ & ~~\phi_{1,2}\phi_{4,1}\\
\!\!\!\phi_{4,1} & \phi_{1,2}\phi_{4,1} & ~~\phi_{4,5}\\
\end{pmatrix}\\[4pt]
\text{down sector:}&~~~~ 
\begin{pmatrix}
\!\!\!\phi_{5,1}\phi_{3,5} & \phi_{5,1}\phi_{3,5} & ~~\phi_{5,1}\phi_{3,5}\\
\!\!\!~~\phi_{5,1}\phi_{3,5}\phi_{1,2}~~ & ~~\phi_{5,1}\phi_{3,5}\phi_{1,2}~~ & ~~\phi_{5,1}\phi_{3,5}\phi_{1,2}\\
\!\!\!\phi_{3,5} & \phi_{3,5} & ~~\phi_{3,5}\\
\end{pmatrix}
\end{aligned}
\end{equation}
with the convention that the ordering of the three families follows the sequence in which they are listed in \eqref{SM_spectrum_intro}. \\[2mm]
As noted above, Yukawa unification is not automatically enforced in string-derived GUT models, which implies that while the structure of singlet insertions in the Yukawa couplings can be described using $SU(5)$ GUT language, the pre-factors, corresponding to order-one coefficients, cannot. In the underlying string context these coefficients depend on the geometrical moduli (complex structure and K\"ahler moduli) of the compactification. They also take into account the effects of a non-trivial K\"ahler potential for the matter fields, as well as contributions from the RG running.\\[2mm]
Before optimising the order-one coefficients, a crucial step that defines the FN paradigm is to assign generic VEVs to the FN scalars such that the observed mass hierarchies and mixing parameters are reproduced qualitatively. For the model at hand, we claim that the following VEV assignments represent a viable choice:
\begin{center}
\begin{tabular}{ccccc}
$\expval{\phi_{5,1}} \sim \epsilon^3$, & $\expval{\phi_{3,5}} \sim \epsilon^4$, & $\expval{\phi_{4,5}} \sim \epsilon$, & $\expval{\phi_{1,2}} \sim \epsilon^5$, & $\expval{\phi_{4,1}} \sim \epsilon^4$~,\\
\end{tabular}
\end{center}
with $\epsilon \approx 0.4$. This choice of VEVs gives the following Yukawa textures:
\begin{equation} \label{eq:yuk_eg_1_intro}
\text{up sector} \sim 
    \begin{pmatrix}
        \epsilon^7 & \epsilon^{12} & \epsilon^4\\
        \epsilon^{12} & \epsilon^{17} & \epsilon^9\\
        \epsilon^4 & \epsilon^9 & \epsilon\\
    \end{pmatrix},
    \quad
\text{down sector} \sim 
    \begin{pmatrix}
        \epsilon^7 & \epsilon^{7} & \epsilon^7\\
        \epsilon^{12} & \epsilon^{12} & \epsilon^{12}\\
        \epsilon^4 & \epsilon^4 & \epsilon^4\\
    \end{pmatrix}
\end{equation} 
leading (once order-one coefficients are included) to the following patterns for the up-quark, down-quark and charged lepton masses:
\begin{equation} \label{eq:yuk_eg_3}
    (m_u, m_c, m_t) \sim(\epsilon^{17},\epsilon^7,\epsilon) ,\quad (m_d, m_s,m_b),~(m_e,m_\mu,m_\tau) \sim (\epsilon^{12},\epsilon^7,\epsilon^4)~.
\end{equation}
Our choice of VEVs as powers of the same parameter $\epsilon$ is inspired by the approach taken in, for example, in Refs.~\cite{Leurer:1992wg, Leurer:1993gy}. Since $\epsilon$ is sufficiently close to $1$, this choice does not introduce a large hierarchy between the various FN scalar VEVs. \\[2mm]
To compare the model against the experimental values for the quark masses, charged lepton masses and CKM mixing parameters, we then optimise the order-one coefficients and the scale $\epsilon$. A modest random search over this $(3\times9+1)$-dimensional parameter space leads to values that are correct within percentage level. \\[2mm]
\noindent We now present our strategy for identifying viable string-inspired FN models. The first step is to specify the horizontal symmetry group $\cG$, for which there are six possible choices. For each of these, we systematically explore all inequivalent charge assignments for the MSSM multiplets. The number of such assignments is small in each case. Next, we specify the allowed number of FN scalar fields, both perturbative (up to $5$) and non-perturbative (up to $4$), that can appear in the spectrum.\\[2mm] 
With these settings fixed, we proceed to search for viable $G$-charge assignments for the FN scalars and corresponding VEV patterns, expressed as integer powers of a single parameter~$\epsilon\lesssim 1$. The range of powers of $\epsilon$ is constrained so that the VEVs of the perturbative singlets satisfy $\expval{\phi}\lesssim 1$ (in Planck units) while those of the non-perturbative singlets satisfy $\expval{\Phi} \ll 1$.  We construct the leading $\cG$-invariant terms contributing to the Yukawa couplings (using the method presented in Appendix~\ref{App:leading_ops}) and check if the resulting Yukawa texture leads to promising values for the quark masses and CKM parameters. At this stage we do not check the masses in the charged lepton sector, as general arguments indicate that this can be achieved with order-one coefficients, provided that the down-quark Yukawa sector is taken care of. \\[2mm] 
Finally, we optimize the order-one coefficients in the quark and charged lepton sectors, along with the parameter $\epsilon$. The goal is to ensure that all measurable quantities align with their experimental values to within a few percent accuracy. 
The order-one coefficients can, in principle, be calculated within string models, as demonstrated in the analysis of Ref.~\cite{Constantin:2024yxh}, which shows that these coefficients can vary by at least an order of magnitude depending on the choice of complex structure. As such, they can contribute to the observed mass and mixing hierarchies. However, in this paper, we adopt the approach of generating the hierarchies directly through powers of FN scalars, excluding cases where hierarchies arise purely from tuning the coefficients. A detailed analysis shows that certain naturalness conditions must be imposed, which in particular constrain the allowed charge assignments for the Standard Model multiplets, as discussed in  Section~\ref{sec:gf}.\\[2mm] 
Even after taking into account such naturalness considerations, the landscape of string FN models is far too large to be searched systematically for successful models, particularly when multiple $U(1)$ symmetries and many scalars are present. For this reason, we resort to heuristic search methods, in particular genetic algorithms (GAs). Another approach would be to use reinforcement learning which proved successful in the search of standard FN models in Ref.~\cite{Harvey:2021oue}). \\[2mm]
So far, we have made no mention of the neutrino sector. The chiral multiplets associated with the FN scalars include fermions that could, in principle, serve as right-handed neutrinos. However, the generation of appropriate neutrino masses (such as through the see-saw mechanism) is more complicated and tied to the problem of moduli stabilization, which complicates the overall dynamics. A detailed treatment of these issues is beyond the scope of this paper and will be explored in a future publication. We emphasize, however, that finding successful FN models that can reproduce the quark and charged lepton sectors has important implications for top-down string model building. The successful $\cG$ charge patterns identified in this way can be translated into, likely strong, topological constraints on string compactifications.\\[2mm] 
It is important to note that the FN mechanism explored here, based on $U(1)$ symmetries from string compactifications, is not the only approach for generating mass and mixing hierarchies in string models. Discrete symmetries can also play a role, both in smooth Calabi-Yau settings and, more extensively, within the framework of modular flavour symmetries (for reviews, see \cite{Feruglio:2017spp}, \cite{deAnda:2018ecu, Feruglio:2019ybq,Kobayashi:2023zzc} and references therein). These modular symmetries can also have a string-theoretic origin \cite{Baur:2021bly,Nilles:2023shk, Ishiguro:2024xph}. \\[2mm]
The rest of the paper is structured as follows. In Section~\ref{sec:susysm} we present the class of four-dimensional $\mathcal N=1$ supergravity theories that underlie the string FN models studied here. We discuss the flavour symmetry $\cG$ and its phenomenological consequences and summarise our strategy for exploring the string FN landscape. After a short review of GAs in Section~\ref{sec:ga} and a discussion of the implementation used for the present purpose, we present our results in Sections~\ref{sec:res_pert} and ~\ref{sec:res_non-pert}. We conclude in Section~\ref{sec:con}.

\section{String Froggatt-Nielsen models} \label{sec:susysm}
In the first part of this section, we describe the general structure of string FN models, within the framework of four-dimensional $\mathcal N=1$ supergravity. We also briefly review the relevant phenomenological constraints and discuss a number of general features of string FN models which will be helpful in the search.

\subsection{Multiplet content of string FN models} \label{sec:mcfn}
Our models have a symmetry $G_{\text{SM}} \cross \cG$, where $G_{\text{SM}} = SU(3) \cross SU(2) \cross U(1)$ is the Standard Model gauge group and $\cG$ contains the additional Green-Schwarz anomalous $U(1)$ factors. Its precise structure is determined by a (non-trivial) partition of five described by an integer vector $\mathbf{n} = \left(n_1,...,n_{f}\right)$, where $f=2,3,4$ or $5$, with all $n_a\geq 1$ and $\sum_{a=1}^{f}n_a=5$. In terms of this partition, the group $\cG$ is given by
\begin{equation} \label{Gdef}
	\cG = \left\{ \left(e^{i\epsilon^1},...,e^{i\epsilon^f}\right) \,\middle\vert\, \epsilon^a\in\mathbb{R}\,,\; \bn\cdot\beps=0  \right\}\cong U(1)^{f-1}\; ,
\end{equation}
where $\beps=(\epsilon^1,\ldots,\epsilon^f)$ are the group parameters and $\cG$ is isomorphic to a product of $f-1$ $U(1)$ factors. These are gauge symmetries but the associated gauge bosons are super-massive and at low energies they appear, effectively, as global symmetries. It is often convenient to describe $\cG$ by using $f$ $U(1)$ symmetries, subject to the constraint in Eq.~\eqref{Gdef}, and we will denote these by  $U_a(1)$, where $a=1,\ldots ,f$. From this viewpoint, representations of $\cG$ can be labelled by integer vectors $\bq=(q_1,\ldots ,q_f)\in\mathbb{Z}^f$, subject to identifying two such vectors $\bq$ and $\bq'$ according to the relation
\begin{equation}\label{equiv}
\bq\sim \bq'\quad\Leftrightarrow\quad
\bq-\bq'\in\mathbb{Z}\bn\; .
\end{equation}
Hence, the $\cG$ charge lattice is given by $\mathbb{Z}^f/\mathbb{Z}{\bf n}$. In particular, this means that the $\cG$-invariant operators are precisely those with charge $\bq\in\mathbb{Z}\bn$.\\[2mm]
The multiplet content of the $\mathcal N=1$ supergravity theory for string FN models consists of gauge multiplets with gauge group $G_{\rm SM}$ (while the $\cG$ vector multiplets are super-massive and have been integrated out) and chiral multiplets. The latter include the standard MSSM multiplets $Q^I,u^I,d^I,L^I,e^I$, where $I=1,2,3$ is a family index, the two Higgs multiplets $H^u,H^d$,  as well as a number $N_\phi+N_\Phi$ of chiral multiplets containing the perturbative FN scalars (bundle moduli) $\phi^{(\alpha)}$, where $\alpha=1,\ldots ,N_\phi$, and the non-perturbative FN scalars (associated with K\"ahler moduli) $\Phi^{(i)}=e^{-T^i}$, where $i=1,\ldots ,N_\Phi$. 

\begin{table}[!ht]
\begin{center}
	\begin{tabular}{|c|c|c||c|c|c|}
		\hline\hline
		field & SM rep & name & SU(5) & $\cG$ charge pattern & SU(5)$\times \cG$ \\[1pt]
		\hline
		$Q$ & ~~$(\bm{3},\bm{2})_{1~~}$ & LH quark&$\bm{10}$ & $\uv_a$ & $\bm{10}_a$\\[2pt]
		$u$ & ~~$(\bm{\bar{3}},\bm{1})_{-4}$& RH $u$-quark & &  &\\[2pt]
		$e$ & ~~$(\bm{1},\bm{1})_{6~~}$& RH electron& &  &\\[2pt]\hline
		$d$ & ~~$(\bm{\bar{3}},\bm{1})_{2~~}$ &RH $d$-quark& $\bm{\bar{5}}$ & $\uv_a+\uv_b$ & $\bm{\bar{5}}_{a,b}$\\[2pt]
		$L$ & ~~$(\bm{1},\bm{2})_{-3}$& LH lepton& & & \\[2pt]\hline
		$H^d$ & ~~$(\bm{1},\bm{2})_{-3}$&down-Higgs& $\bm{\bar{5}}^{H^d}$ & $\uv_a+\uv_b$ & $\bm{\bar{5}}_{a,b}^{H^d}$\\[2pt]
		$H^u$ & ~~$(\bm{1},\bm{2})_{3~~}$&up-Higgs& $\bm{5}^{H^u}$ & \!\!\!\!$-\uv_a-\uv_b$ & $\bm{5}_{a,b}^{H^u}$\\[2pt]\hline
		$\phi$ & $(\bm{1},\bm{1})_0$&pert.~FN scalar& $\bm{1}$ & $\uv_a-\uv_b$ & $\mathbf{1}_{a,b}$\\[2pt]
		$\Phi$ & $(\bm{1},\bm{1})_0$&non-pert.~FN scalar & $\bm{1}$ & $\bk=(k_1,\ldots ,k_f)$ & $\mathbf{1}$\\[2pt]
		\hline
	\end{tabular}
\caption{The field content and multiplet charges of string FN models. In addition to the $G_{\rm SM}$ representations, we have also included the associated $SU(5)$ GUT representations for the usual embedding $G_{\rm SM}\subset SU(5)$. 
The fifth column provides the $\cG$ charge pattern, where $\{\uv_a\}$ are the $f$-dimensional standard basis vectors. The indices $a$ and $b$ indicate under which $U(1)$ symmetries multiplets carry charge and depend on the family.  
For multiplets with charge $\pm(\uv_a+\uv_b)$, the choice $a=b$ is permitted only if the partition vector $\mathbf{n}$ satisfies $n_a>1$. The $\cG$ charges for the non-perturbative FN~scalars~$\Phi^{(i)}$ are integer vectors $\bk^i=({k^i}_a)\in\mathbb{Z}^f$ satisfying $\sum_{a=1}^f{k^i}_a=0$. }\label{tab:fields} 
\end{center}
\end{table}

\noindent The list of fields and their representations are given explicitly in Table~\ref{tab:fields}, where we have not only provided the $G_{\rm SM}$ representations but also the $SU(5)$ GUT representations for the usual embedding $G_{\rm SM}\subset SU(5)$. The appearance of  the $SU(5)$ multiplet structure is related to the structure of the underlying string models. 
The group $\cG$ can be viewed as a collection of $U(1)$ flavour symmetries with a specific pattern of charges indicated in the fifth column in Table~\ref{tab:fields}, which depends only on the $SU(5)$ representation. For this reason, it will be convenient to use $SU(5)$ GUT notation for the fields and operators. However, this does not mean that we consider GUTs -- our models are MSSM-like with the Standard Model gauge group\footnote{\noindent These models are based on rank five vector bundles and, initially, lead to $SU(5)$ GUT models. However, the $SU(5)$ group is broken to the Standard Model group by a Wilson-line bundle and the GUT symmetry is never realised at the level of four-dimensional field theories.}. While the charges and, hence, the singlet insertions follow an $SU(5)$ GUT pattern this is not, in general the case for the coefficients in front of these insertions. This means we have a crude form of Yukawa unification at the level of allowed singlet insertions which is generically broken by the coefficients multiplying these insertions~\cite{Buchbinder:2016jqr}.\\[2mm]
Any of the Standard Model multiplets and the perturbative FN scalars from Table~\ref{tab:fields}, denoted generically by $C$ and having $\cG$ charge $\bq(C)$, transform linearly as
\begin{equation}
 C\mapsto e^{-i\bq(C)\cdot\beps}C\; .
\end{equation}
On the other hand, the K\"ahler moduli $T^i$ with $\cG$ charge $\bk^i\in\mathbb{Z}^f$ transform non-linearly as:
\begin{equation} \label{eq:kahl_tf}
  T^i\mapsto T^i+i\bk^i\cdot \beps\qquad\Leftrightarrow\qquad
  \Phi^{(i)}=e^{-T^i}\mapsto e^{-i\bk^i\cdot\beps}\,\Phi^{(i)}\; ,
\end{equation}
where the charge vectors $\bk^i=({k^i}_a)$ are subject to the constraint
\begin{equation}\label{c10}
     \sum_{a=1}^f{k^i}_a=0~,\qquad\forall \;i\in\{1,\ldots ,N_{\Phi}\}\; .
\end{equation}
(In a string context, this condition corresponds to the vanishing first Chern class of the underlying split vector bundle.)

\subsection{Anomaly cancellation}
As mentioned above, the $U(1)$ symmetries in $\cG$ are typically anomalous. In particular, the mixed triangle anomalies of the form $U_a(1)\times SU(3)^2$, $U_a(1)\times SU(2)^2$, $U_a(1)\times U(1)^2$ which involve one $U_a(1)$ and two Standard Model gauge bosons, are non-vanishing. For the three factors in $G_{\rm SM}$ these anomalies are proportional to
\begin{equation}\label{tanomalies}
 \begin{array}{rcl}
 \mathbf{A}^{(3)}&=&\mathbf{A}^{(5)}\\
\mathbf{A}^{(2)}&=&\mathbf{A}^{(5)}+\bq(H^u)+\bq(H^d)\\
\mathbf{A}^{(1)}&=&\mathbf{A}^{(5)}+\frac{3}{5}\left(\bq(H^u)+\bq(H^d)\right)
\end{array}
\end{equation} 
where $\mathbf{A}^{(5)}$ is the mixed GUT $U_a(1)SU(5)^2$ anomaly, 
\begin{equation}
    \mathbf{A}^{(5)}=\sum_{I=1}^3\left(\bq(\bar{\bf 5}^I)+3\,\bq({\bf 10}^I)\right)\; ,
\end{equation}
evaluated only on the families $\bar{\bf 5}^I$ and ${\bf 10}^I$ (but not on the Higgs multiplets which are incomplete from an $SU(5)$ GUT perspective).\\[2mm]
These triangle anomalies have to be cancelled by the Green-Schwarz mechanism, facilitated by a non-trivial $\cG$ transformation of the gauge kinetic function for the $G_{\rm SM}$ gauge bosons. In principle, this gauge kinetic function can be different for the three factors in~$G_{\rm SM}$. However, for models derived from heterotic compactifications, the gauge kinetic function is universal and we adopt this feature for our string FN models. From Eq.~\eqref{tanomalies}, this means we require, in a first instance, that the two Higgs doublets have opposite $\cG$-charge, that is, $\bq(H^u)+\bq(H^d)\sim {\bf 0}$, so that all three anomaly coefficients are the same. The universal gauge kinetic function for $G_{\rm SM}$ is given by
\begin{equation}
 f=S+\beta_iT^i\; ,
\end{equation}
where $\beta_i\in\mathbb{Z}$ are integers and $S$ is the dilaton multiplet with $\cG$ transformation
\begin{equation}
 S\mapsto S+i\beta_i\beps\cdot\bk^i\; .
\end{equation}
For Green-Schwarz anomaly cancellation to work, we require that
\begin{equation} \label{anomalyc}
 \mathbf{A}^{(5)}\sim\beta_i\bk^i\; ,
\end{equation}
and this condition will be imposed on our models. 

\subsection{SUGRA Lagrangian of string FN models} \label{sec:sugra_lang}
The $R$-parity conserving\footnote{Here we assume the existence of a symmetry, such as the standard R-Parity or a matter parity, which originates from the presumed underlying string model and forbids R-parity violating terms.} K\"ahler potential and superpotential for our models are given by 
\begin{eqnarray} 
K &=& \hat{k}^u H^u \bar{H}^u + \hat{k}^d H^d \bar{H}^d + \hat{K}^Q_{IJ} Q^I \bar{Q}^J + \hat{K}^u_{IJ} u^I \bar{u}^J + \hat{K}^d_{IJ} d^I \bar{d}^J  \\ 
&& ~~~~~~~~+ \hat{K}^{L}_{IJ}L^I\bar{L}^J + \hat{K}^{e}_{IJ}e^I\bar{e}^J + \cdots \label{K}\\[2mm]
W &=& \hat{\Lambda}^u_{IJ} H^u Q^I u^J + \hat{\Lambda}^d_{IJ} H^d Q^I d^J + \hat{\Lambda}^e_{IJ} H^d L^I e^J + \cdots~, \label{W} \end{eqnarray}
where the dots stand for terms that involve only the chiral multiplets in which the FN scalars live. %
The K\"ahler potential for the gravitational moduli (which exhibits a no-scale structure at tree level) has been omitted as it only depends on moduli quantities uncharged under~$\cG$, and hence places no further constraints relevant to our discussion.
All hatted quantities are coupling functions which contain the possible singlet insertions in terms of $\phi^{(\alpha)}$ and $\Phi^{(i)}$ which lead to $\cG$-invariant operators. More concretely, the Yukawa coupling functions are schematically given by
\begin{equation}\label{Lhat}
    \hat{\Lambda}^d_{IJ}=\sum_r a^d_{IJ,r}{\cal O}^d_{IJ,r}\;,\qquad
    \hat{\Lambda}^u_{IJ}=\sum_r a^u_{IJ,r}{\cal O}^u_{IJ,r}\; ,\qquad
    \hat{\Lambda}^e_{IJ}=\sum_r a^e_{IJ,r}{\cal O}^e_{IJ,r}\; ,
\end{equation}
where $a^d_{IJ,r}$ and $a^u_{IJ,r}$ are order-one coefficients which are not fixed by any of the symmetries under consideration. The terms ${\cal O}^d_{IJ,r}$, ${\cal O}^u_{IJ,r}$, ${\cal O}^e_{IJ,r}$ are $\phi^{(\alpha)}$ and $\Phi^{(i)}$ singlet insertions (counted by the index $r$) which lead to $\cG$-invariant operators, that is, they satisfy
\begin{equation}
\begin{aligned}
\bq({\cal O}^u_{IJ,r})\sim -&\bq(H^u Q^Iu^J)\;,~~
\bq({\cal O}^d_{IJ,r})\sim -\bq(H^d Q^Id^J)\;,~~\\[2mm]
&~~~\bq({\cal O}^e_{IJ,r})\sim -\bq(H^d L^I e^J)\; .
\end{aligned}
\end{equation}
The matter field metrics in the K\"ahler potential are constructed in a similar way, for example
\begin{equation}\label{KQ}
  K^Q_{IJ}=b^Q_I\delta_{IJ}+\sum_r b^Q_{IJ,r}{\cal O}^Q_{IJ,r}
\end{equation}
Here, $b_I^Q$ and $b_{IJ,r}^Q$  are order-one coefficients~\footnote{The overall scale for the matter field K\"ahler potential $\sim (T + \bar{T})^{-1}$ can be accounted for by an appropriate redefinition of the order-one coefficients as well as the reference scale for the VEVs of the singlet fields. To simplify the discussion, we have absorbed the moduli-dependence into the order-one coefficients and definition of the reference scales. Ultimately, deciding whether the moduli-dependent coefficients $b^Q_I$ and $b^Q_{IJ,r}$ are indeed of order one in any given compactification, can be accomplished by performing numerical calculations of the matter field K\"ahler metric analogous to those carried out in Ref.~\cite{Constantin:2024yxh}.} and the singlet insertions ${\cal O}^Q_{IJ,r}$ are monomials in the singlet fields (and their complex conjugates) with
\begin{equation}
 \bq\left({\cal O}^Q_{IJ,r}\right)\sim -\bq\left(Q^I\bar{Q}^J\right)\; .
\end{equation}
Note that $\bq\left(Q^I\bar{Q}^I\right)\sim{\bf 0}$ so that the diagonal kinetic terms are always allowed without any singlet insertion, as indicated by the first term on the RHS of Eq.~\eqref{KQ}. Similar expressions and statements apply to the other field space metrics in Eq.~\eqref{K}.\\[2mm] 
In order to illustrate how to find the allowed singlet insertions, we consider an explicit example with split pattern $\bn=(1,1,1,1,1)$ and a multiplet content which includes $\bar{\bf 5}_{1,2}$, ${\bf 10}_2$, ${\bf 10}_5$, $\bar{\bf 5}_{4,5}^H$ and ${\bf 5}^H_{4,5}$ (using GUT language for simplicity). Then, the up-Yukawa term ${\bf 5}^H_{4,5}{\bf 10}_5{\bf 10}_5$ carries $\cG$-charge $\bq\left({\bf 5}^H_{4,5}{\bf 10}_5{\bf 10}_5\right)=(0,0,0,-1,1)$. This means possible singlet insertions are, for example, $\phi_{4,5}$ or $\phi_{4,1}\phi_{1,5}$ (provided singlets with these charges appear in the theory). On the other hand, the down-Yukawa coupling $\bar{\bf 5}_{4,5}^H\bar{\bf 5}_{1,2}{\bf 10}_5$ has $\cG$-charge $\bq\left(\bar{\bf 5}_{4,5}^H\bar{\bf 5}_{1,2}{\bf 10}_5\right)=(1,1,0,1,2)$ so that a singlet insertion  $\phi_{3,5}$ is possible. Note, this leads to $\bq\left(\phi_{3,5}\bar{\bf 5}_{4,5}^H\bar{\bf 5}_{1,2}{\bf 10}_5\right)=(1,1,1,1,1)$ and, from Eq.~\eqref{equiv}, this charge is indeed equivalent to charge ${\bf 0}$. In summary, we are following standard FN lore, but with multiple $U(1)$ symmetries and scalar fields and specific charge patterns, as suggested by string theory.

\subsection{Masses and mixing} \label{sec:mass_mix}
After having written down the most general $\cG$-invariant theory (up to a certain maximal operator dimension), as explained in the previous sub-section, we consider singlet VEVs $\langle\phi^{(\alpha)}\rangle<1$ and $\langle\Phi^{(i)}\rangle\ll 1$, thereby converting the various coupling functions into coupling constants
\begin{equation} K^u=\hat{K}^u\left(\langle\phi^{(\alpha)}\rangle,\langle\Phi^{(i)}\rangle\right)\;,\quad 
\Lambda_{IJ}^u=\hat{\Lambda}_{IJ}^u\left(\langle\phi^{(\alpha)}\rangle,\langle\Phi^{(i)}\rangle\right)\;,\quad
\dots
\end{equation} 
and similarly for the other couplings. To work out quark masses and mixing we should first canonically normalise kinetic terms, that is, we should introduce ${\rm GL}(3,\mathbb{C})$ matrices $P^Q$, $P^u$ and $P^d$ such that
\begin{equation}
 (P^Q)^\dagger K^QP^Q=\bm{1}_3\;,\quad
 (P^u)^\dagger K^uP^u=\bm{1}_3\;,\quad
 (P^d)^\dagger K^dP^d=\bm{1}_3\;.
\end{equation}
Then, the `physical' mass matrices $M^u$ and $M^d$, relative to canonically normalised kinetic terms, are given by
\begin{equation}\label{M0}
 M^u=\frac{v^u}{\sqrt{k^u}}(P^Q)^\dagger\Lambda^u P^u\;,\qquad
 M^d=\frac{v^d}{\sqrt{k^d}}(P^Q)^\dagger\Lambda^d P^d\; ,
\end{equation}
where $v^u$ and $v^d$ are the up- and down-Higgs VEVs, respectively, satisfying
\begin{equation}
 \sqrt{(v^u)^2+(v^d)^2}=v\simeq 174\;{\rm GeV}\; ,
\end{equation}
with the electro-weak VEV $v$. The discussion can be somewhat simplified owing to the structure of the matter field metrics as, for example, given in Eq.~\eqref{KQ}. The order-one terms along the diagonal dominate over the additional singlet insertions, so that the matrices $P^Q$, $P^u$ and $P^d$ are approximately diagonal with order-one diagonal entries. Rotating with these matrices amounts to another order-one effect which we can think of as being absorbed by the order-one coefficients in $\hat{\Lambda}^u$ and $\hat{\Lambda}^d$. Therefore, in practice, since we are only concerned about the leading-order terms in the singlet fields, we only need to construct the singlet insertions for the Yukawa functions $\hat{\Lambda}_{IJ}^d$ and $\hat{\Lambda}_{IJ}^u$ and can omit the sub-leading order effects from canonically normalising the kinetic terms. Similar considerations apply to the charged leptons and the Higgs fields. \\[2mm]
It follows that we can simply write
\begin{equation}
    M^u=v^u\Lambda^u\;,\qquad M^d=v^d\Lambda^d\;,\qquad M^e=v^d\Lambda^e\; .
\end{equation}
As usual, these mass matrices need to be diagonalised to obtain the quark and charged lepton masses so we introduce unitary matrices $U^u$, $V^u$, $U^d$, $V^d$, $U^e$, $V^e$ such that
\begin{equation}\label{diag}
\begin{array}{rclcrcl}
 M^u &=& U^u \hat{M}^u (V^u)^\dagger&\qquad&\hat{M}^u &=& \text{diag}(m_u,m_c,m_t)\\
 M^d &=& U^d \hat{M}^d (V^d)^\dagger&&\hat{M}^d &=& \text{diag}(m_d,m_s,m_b)\\
  M^e &=& U^e \hat{M}^e (V^e)^\dagger&&\hat{M}^e &=& \text{diag}(m_e,m_\mu,m_\tau)\; .
 \end{array}
\end{equation} 
The Cabibbo-Kobayashi-Maskawa (CKM) quark mixing matrix $V_{\rm CKM}$ is given by
\begin{equation}\label{VCKM}
        V_{\rm CKM} = (U^u)^\dagger U^d~.
\end{equation}
The above Yukawa couplings are those just below the compactification scale and, in principle, they have to be RG evolved down to low energies for a comparison with measured masses and mixing. It is well-known that this leads to order-one rescalings of the Yukawa couplings, and these will be taken into account using the approximate solution to the RG equations described in Appendix~\ref{sec:rg}. \\[2mm]
For later reference, the measured masses for the quarks and charged leptons are listed in Table~\ref{tab:exp}.
\begin{table}[!h]
\renewcommand{\arraystretch}{1.35}
\centering
\begin{tabular}{|c|c|c|c|}
\hline
\textbf{Quark} & \(m_u\) (GeV) & \(m_c\) (GeV) & \(m_t\) (GeV) \\
\hline
\textbf{Mass} & \(0.00216^{+0.00049}_{-0.00026}\) & \(1.27 \pm 0.02\) & \(172.4 \pm 0.07\) \\
\hline
\textbf{Quark} & \(m_d\) (GeV) & \(m_s\) (GeV) & \(m_b\) (GeV) \\
\hline
\textbf{Mass} & \(0.00467^{+0.00048}_{-0.00017}\) & \(0.093^{+0.011}_{-0.005}\) & \(4.18^{+0.03}_{-0.02}\) \\
\hline\hline
\textbf{Lepton} & \(m_e\) (GeV) & \(m_\mu\) (GeV) & \(m_\tau\) (GeV) \\
\hline
\textbf{Mass} & \(0.000511 \pm 0.000001\) & \(0.1057 \pm 0.0001\) & \(1.77682 \pm 0.00016\) \\
\hline
\end{tabular}
\caption{\sf Measured masses of quarks and charged leptons in GeV, as reported in Ref.~\cite{ParticleDataGroup:2024cfk}.}
\label{tab:exp}
\end{table}

\noindent The CKM matrix, taken from Ref.~\cite{ParticleDataGroup:2024cfk}, is given by
\begin{equation}
        \label{eqn:CKMVal}
        \left| V_{CKM} \right| \simeq \left(
            \begin{array}{ccc}
             0.97373 \pm 0.00031 & ~~0.2243 \pm 0.0008~~ & 0.00382 \pm 0.00020 \\[2pt]
             0.221 \pm 0.004 & 0.975 \pm 0.006 & 0.0408 \pm 0.0014 \\[2pt]
             0.0086 \pm 0.0002 & 0.0415 \pm 0.0009 & 1.014 \pm 0.029 \\
            \end{array}
            \right)\; .
\end{equation}

\section{The landscape of string FN models}\label{sec:gf}
\subsection{A first estimate of the size}
Given a horizontal symmetry $\cG\cong U(1)^{f-1}$, associated with a partition $\mathbf{n}$ of $5$ with $f$ parts, we estimate the number of charge assignments for the SM multiplets and the FN singlets.\\[2mm]  
For the three $\mathbf{10}_a$ multiplets, there are $f^3$ choices. For each $\bar{\bf 5}_{a,b}$ or ${\bf 5}_{a,b}$ multiplet there are, roughly speaking, $f^2/2$ choices (depending on the partition $\mathbf n$, charges with $a=b$ may or may not be allowed), giving a total of $f^8/16$ choices. Recall that the two Higgs multiplets come in a vector-like pair, so their $\cG$ charges are correlated. For each perturbative singlet~$\phi^{(\alpha)}$ there are $f^2$ possibilities, giving a total of $f^{2N_\phi}$ choices. For each non-perturbative singlet~$\Phi^{(i)}$, taking into account the constraint~\eqref{c10}, there are $\Delta k^{f-1}$ possibilities, giving a total of $\Delta k^{(f-1)N_\Phi}$ choices, where we assume a range of charges ${k^i}_a\in\{k_{\rm min},\ldots ,k_{\rm max}\}$ and $\Delta k=k_{\rm max}-k_{\rm min}+1$.\\[2mm]
Combining all this gives an approximate total number of charge choices of
\begin{equation}\label{Ncharges0}
    N_{\rm charges}\simeq \frac{f^{11+2N_\phi}\Delta k^{(f-1)N_\Phi}}{16}\; .%{2^{N_\phi+4}}\; .
\end{equation}
For instance, for $f=5$, $\Delta k=10$, $N_\phi=4$ and $N_\Phi=2$ this leads to approximately $N_{\rm charges}\simeq 10^{20}$ choices. 
However, not all of these configurations are physically distinct. The $\mathbf{10}$ and $\bar{\mathbf{5}}$ families can be permuted under an $S_3 \times S_3$ symmetry, while the singlet fields can be permuted under an $S_{N_\phi} \times S_{N_\Phi}$ symmetry. Additionally, there is a permutation symmetry for the $U_a(1)$ factors, governed by the subgroup of $S_f$ that stabilizes the vector~$\mathbf{n}$. These symmetries reduce the number of inequivalent charge assignments by the order of the combined symmetry group. In the above example, this gives $\simeq 10^{15}$ inequivalent charge assignments.
However, for the heuristic search method described below it will not be practical to eliminate the permutation symmetry so we will be dealing effectively with the number in Eq.~\eqref{Ncharges0}.\\[2mm]
The choice of singlet VEVs also needs to be included in the estimation of the total size of the string FN landscape. The reference scale for the singlet VEVs is the four-dimensional Planck scale which, in a string context, can be verified by a straightforward reduction calculation from ten dimensions~\footnote{Together with an appropriate rescaling to account for overall factors in the K\"ahler potential and superpotential due to canonically-normalising the kinetic terms, as discussed in \S\ref{sec:mass_mix}.}. The VEVs can, of course,  take values in a continuous range but, in the spirit of FN model building, they will be discretised by writing
\begin{equation}\label{VEVs}
 \langle\phi^{(\alpha)}\rangle= \epsilon^{p_\alpha}
\;,\qquad
\langle \Phi^{(i)}\rangle =\epsilon^{\tilde p_i}\; ,
\end{equation}
where $0<\epsilon<1$ is a model-dependent number which we refer to as the `$\epsilon$-scale' and $p_\alpha \in\{1,\ldots ,p_{\rm max}\}$, $\tilde p_i\in\{\tilde p_{\rm min},\ldots ,\tilde p_{\rm max}\}$ are non-negative integers. Since the VEVs for the non-perturbative singlets $\Phi^{(i)}$ are exponentially suppressed, their typical values will be taken to be smaller than the $\phi^{(\alpha)}$ VEVs.
This will be implemented by different ranges for the integers $p_\alpha$ and $\tilde p_i$, setting $\tilde{p}_{\text{min}} > 1$. The total number of models in the string FN landscape, including the number of VEV choices, is
\begin{equation}
    N_{\rm models}=N_{\rm charges}\;\Delta p^{N_\phi} \Delta \tilde p^{N_\Phi}\; ,
\end{equation}
where $\Delta p=p_{\rm max}$ and $\Delta \tilde p=\tilde p_{\rm max}-\tilde p_{\rm min}+1$. For the example $f=5$, $\Delta k=10$, $N_\phi=4$ and $N_\Phi=2$ and choosing $\Delta p=\Delta \tilde p=10$ this leads to, roughly, $N_{\rm models}\simeq 10^{26}$ models. It's therefore clear that, at least for the cases with many singlet fields, the landscape of string FN models is large enough that heuristic search algorithms are a sensible strategy.

\subsection{Naturalness conditions}\label{sec:naturalness}
Another aspect of FN model building is that family mass hierarchies should result from either charge choices or different VEVs (or a combination of both), but not from fine-tuning of the ${\cal O}(1)$ coefficients in Eq.~\eqref{Lhat}. In order to formulate such a non-fine-tuning condition, consider a mass matrix $M$ ($M^u$ or $M^d$ in our case) and define $\cM=MM^\dagger$. From our parametrisation of VEVs in Eq.~\eqref{VEVs} and controlled by the $\cG$-symmetry and charge choices, the entries of $\cM$ are organised by powers of $\epsilon$, that is,
\begin{equation}\label{Meps}
     \cM_{IJ}=c_{IJ}\epsilon^{\nu_{IJ}}\; ,
\end{equation}
where $|c_{IJ}|={\cal O}(1)$ with $c_{JI}=c_{IJ}^*$ and $\nu_{IJ}=\nu_{JI}\in \{0,1,\ldots \}$. For illustration, consider a simple two-family example with matrix
\begin{equation}
 \cM=\left(\begin{array}{cc}c_{11}\epsilon^{\nu_{11}}&c_{12}\epsilon^{\nu_{12}}\\c_{12}^*\epsilon^{\nu_{12}}&c_{22}\epsilon^{\nu_{22}}\end{array}\right)\; .
\end{equation}
We want to require that the two eigenvalues of $\cM$ are proportional to different powers of $\epsilon$. It is easy to show, by considering ${\rm tr}(\cM)$ and ${\rm det}(\cM)$, that this is equivalent to the `naturalness condition'
\begin{equation}\label{nucond}
    \left.\begin{array}{l}\cM\mbox{ eigenvalues with}\\\mbox{different powers of }\epsilon\end{array}\right\}
    \quad\Leftrightarrow\quad
    {\rm min}\left(\frac{\nu_{11}+\nu_{22}}{2},\nu_{12}\right)>{\rm min}(\nu_{11},\nu_{22})\; .
\end{equation}
For example, if $\nu_{11}=\nu_{22}$ this condition cannot be satisfied so, in this case, the eigenvalues of $\cM$ are given by the same power of $\epsilon$. Then, the only way to obtain eigenvalues with a numerical hierarchy is by tuning the coefficients $c_{IJ}$ -- and this is precisely what we would like to exclude. For this reason, we impose the condition~\eqref{nucond} on the powers $\nu_{IJ}$ or, rather, we impose its three-family analogue which can be easily derived by considering
\begin{eqnarray}
\lambda_3+\lambda_2+\lambda_1&=&{\rm tr}(\cM)={\cal O}(\epsilon^{r_3})\\
\lambda_3\lambda_2+\lambda_3\lambda_1+\lambda_2\lambda_1&=&\frac{1}{2}\left({\rm tr}(\cM)^2-{\rm tr}(\cM^2)\right)={\cal O}(\epsilon^{r_2})\label{eq:nc_quad}\\ \lambda_1\lambda_2\lambda_3&=&{\rm det}(\cM)={\cal O}(\epsilon^{r_1})\; ,
\end{eqnarray}
where $\lambda_1\ll\lambda_2\ll\lambda_3$ are the eigenvalues of $\cM$ and the $r_I$ specify the leading $\epsilon$ powers of each of the three invariants, computed directly from the matrix~\eqref{Meps}. Clearly, $\lambda_3={\cal O}(\epsilon^{r_3})$, $\lambda_2={\cal O}(\epsilon^{r_2-r_3})$ and $\lambda_1={\cal O}(\epsilon^{r_1-r_2-r_3})$ so, in order to have three eigenvalues with different orders in $\epsilon$, we require the naturalness conditions
\begin{equation} \label{eq:nat_cont}
 2\, r_3<r_2\;,\qquad 2\,r_2<r_1\; .
\end{equation}
For example, consider the matrix
\begin{equation}
\cM={\cal O}
\begin{pmatrix}
\epsilon^5 & \epsilon^3 & \epsilon^2 \\
\epsilon^3 & \epsilon^2 & \epsilon \\
\epsilon^2 & \epsilon & 1 \\
\end{pmatrix}
\end{equation}
for which $r_3=0$, $r_2=2$ and $r_1=6$. Then $\lambda_3={\cal O}(1)$, $\lambda_2={\cal O}(\epsilon^2)$ and $\lambda_1={\cal O}(\epsilon^4)$, so we have three eigenvalues with an $\epsilon$ hierarchy and the model passes the naturalness conditions. On the other hand, for
%page 30
\begin{equation}
\cM={\cal O}
\begin{pmatrix}
\epsilon^3 & \epsilon^2 & \epsilon \\
\epsilon^2 & \epsilon^2 & \epsilon \\
\epsilon^2 & \epsilon & 1 \\
\end{pmatrix}
\end{equation}
it follows that $r_3=0$, $r_2=2$ and $r_1=4$. Hence, $\lambda_3={\cal O}(1)$, $\lambda_2={\cal O}(\epsilon^2)$ and $\lambda_1={\cal O}(\epsilon^2)$, so this model fails the naturalness test.\\[2mm]
The above naturalness test significantly constraints the range of viable charge patterns. A detailed analysis, summarised in Appendix \ref{sec:nc_qfix}, leads to the following two necessary conditions:
\begin{itemize}
\item The spectrum should contain no pair of $\bm{10}$-multiplets with identical $\cG$-charge.
\item If $\bar{\bf 5}^H_{a,b}$ is the multiplet containing the down-Higgs, the spectrum cannot contain both a ${\bf 10}_a$ and a ${\bf 10}_b$ multiplet.
\end{itemize}
Combining these constraints with the $S_3$ permutation symmetry in the ${\bf 10}$-sector leads to a small number of possible charge choices for the ${\bf 10}$ multiplets and the $\bar{\bf 5}^H$ multiplet which are listed in Table~\ref{tab:10}. These charge choices are selected such that any of the other charge choices for a fixed vector $\mathbf{n}$ will be equivalent to one of them up to the redundant permutation symmetry $S_3\times S_3 \times S_{N_\phi} \times S_{N_\Phi} \times S_f$. In picking these charge choices, the redundant symmetry is broken down into a smaller subgroup as described in Table~\ref{tab:10} for each of the charge choices. It is notable that the split patterns $\bn=(1,4)^T$ and $\bn=(2,3)^T$ are ruled out completely. For the remaining cases with $|\bn| \geq 3$, a short calculation shows that in order to generate the hierarchy in the up-Yukawa sector, at least two FN singlets are required. 

\begin{table}[ht]
\begin{center}
	\begin{tabular}{|c|c|c|c|c|c|c|c|}
	\hline
	$\mathbf{n}$ & \multicolumn{4}{c|}{Charges} & Symmetry after $q$-fixing & $|G_{\rm red}|$\\[1pt]
	\hline
	$(1,1,1,1,1)$ & $\bm{10}_1$ & $\bm{10}_2$ & $\bm{10}_5$ & $\bm{\bar{5}}^H_{4,5}$ & $S_3 \times S_{N_{\phi}} \times S_{N_\Phi}\times S_2$ & $12N_\phi!N_\Phi!$\\[1pt]
	\hline
	\multirow{4}{*}{$(1,1,1,2)$} & $\bm{10}_1$ & $\bm{10}_2$ & $\bm{10}_4$ & $\bm{\bar{5}}^H_{4,4}$ & $S_3 \times S_{N_{\phi}} \times S_{N_\Phi}\times S_2$ & $12N_\phi!N_\Phi!$\\[1pt]
	 & $\bm{10}_2$ & $\bm{10}_3$ & $\bm{10}_4$ & $\bm{\bar{5}}^H_{1,4}$ & $S_3 \times S_{N_{\phi}} \times S_{N_\Phi}\times S_2$ & $12N_\phi!N_\Phi!$\\[1pt]
	 & $\bm{10}_1$ & $\bm{10}_2$ & $\bm{10}_3$ & $\bm{\bar{5}}^H_{1,4}$ & $S_3 \times S_{N_{\phi}} \times S_{N_\Phi}\times S_2$ & $12N_\phi!N_\Phi!$\\[1pt]
	 & $\bm{10}_1$ & $\bm{10}_3$ & $\bm{10}_4$ & $\bm{\bar{5}}^H_{1,2}$ & $S_3 \times S_{N_{\phi}} \times S_{N_\Phi}$ & $6N_\phi!N_\Phi!$\\[1pt]
	\hline
	$(1,1,3)$ & $\bm{10}_1$ & $\bm{10}_2$ & $\bm{10}_3$ & $\bm{\bar{5}}^H_{3,3}$ & $S_3 \times S_{N_{\phi}} \times S_{N_\Phi}\times S_2$ & $12N_\phi!N_\Phi!$\\[1pt]
	\hline
	$(1,2,2)$ & $\bm{10}_1$ & $\bm{10}_2$ & $\bm{10}_3$ & $\bm{\bar{5}}^H_{3,3}$ & $S_3 \times S_{N_{\phi}} \times S_{N_\Phi}$& $6N_\phi!N_\Phi!$\\[1pt]
	\hline
	\end{tabular}
 \caption{List of charge choices for ${\bf 10}^I$ and $\bar{\bf 5}^H$ required for a non-fine-tuned model. For the last two split patterns, $\bn=(1,4)$ and $\bn=(2,3)$, no viable charge choices remain. The third column lists the order of the permutation symmetry $G_{\rm red}$ left in each sector.}\label{tab:10}
\end{center}
\end{table}

\noindent For each of the $7$ rows in Table~\ref{tab:10}, the remaining number of charge assignments and the overall number of model, including VEV choices is given by
\begin{equation}\label{Ncharges}
    \tilde{N}_{\rm charges}\simeq \frac{f^{6+2N_\phi}\Delta k^{(f-1)N_\Phi}}{8}\;, \qquad
    \tilde{N}_{\rm models}\simeq \tilde{N}_{\rm charges}\Delta p^{N_\phi}\Delta \tilde p^{N_\Phi}\; .
\end{equation}
For the example, with $f=5$, $\Delta k=10$, $N_\phi=4$, $N_\Phi=2$ and $\Delta p=\Delta \tilde p=10$, this leads to $\tilde{N}_{\rm models}\simeq 10^{23}$, still a very large landscape of string FN models.\\[2mm]
The actual number of physically distinct models is obtained by dividing the total number of models $\tilde{N}_{\text{models}}$ in Eq.~\eqref{Ncharges} by the order of the remaining permutation symmetry. For cases with a number of FN singlets less than four, this reduction may lead to a reasonably small environment size which facilitates a complete scan of the model space.  
For example, for the case $f = 3$, $\Delta k = 8$, $N_{\phi} = 3$, $N_\Phi = 0$, $\Delta p = \Delta \tilde p = 8$, leads to $\tilde{N}_{\rm models}/|G_{\text{red}}|\simeq 10^{6}$ models after removing symmetry redundancies, so that a complete scan is computationally feasible. On the other hand, the choice $f=5$, $\Delta k=10$, $N_\phi=4$, $N_\Phi=2$ and $\Delta p=\Delta \tilde p =10$ leads to $\tilde{N}_{\rm models}/|G_{\text{red}}|\simeq 10^{21}$ models and heuristic search methods have to be used.

\subsection{The search strategy} \label{sec:strat_search}
Our approach to exploring the string FN landscape has two main phases. First, we perform a GA search (or, for cases with a sufficiently small number of models, a systematic search) to identify charge assignments for the SM multiplets and FN singlets, together with $\epsilon$-powers for the singlet VEVs that generate promising Yukawa textures which lead to roughly correct quark mass hierarchies and mixing. 
The Yukawa textures are obtained by constructing the leading $\cG$-invariant operators using the method presented in Appendix~\ref{App:leading_ops}.
This step focuses on the quark sector only, as the charged lepton mass hierarchy approximately follows the down-quark sector. \\[2mm]
In the second phase, we optimize the order-one coefficients in all three Yukawa matrices as well as the $\epsilon$-scale, ensuring the Higgs VEV, quark and lepton masses, and CKM matrix match their experimentally measured values at the electroweak scale $M_Z$ within a few percent. A few further remarks are in order.\\[2mm]
The string FN models studied here involve two types of singlets, perturbative and non-perturbative, each coming with a distinct pattern of $\cG$-charges.
For any model containing only perturbative singlet fields, it is straightforward to construct (typically many) physically equivalent models by enhancing the spectrum with additional non-perturbative singlet fields whose contributions to the Yukawa matrices are sub-leading.\\[2mm]
The converse direction is not entirely straightforward and, in general, it is not possible to associate to any given model including both perturbative and non-perturbative singlets a physically equivalent model that is entirely perturbative. The charges of non-perturbative singlet fields can, in principle, be reproduced by suitable combinations of perturbative singlets. However, the VEVs of the charge-equivalent combinations of perturbative singlets must also match the VEV of the non-perturbative singlet. This may not be compatible with the allowed VEV ranges and, moreover, the new operator insertion obtained by replacing a non-perturbative singlet may not remain leading-order. An example illustrating the relation between perturbative and non-perturbative models in given in Section~\ref{sec:ex2_nonpert}.\\[2mm]
Finally, we note that models with a smaller number of singlets can be included as special cases of models with a higher number of singlets (perturbative, non-perturbative or both). This can be done either by adding perturbative singlets of the form $\phi_{a,a}$ which are $\cG$-neutral and hence do not contribute to the leading order operators or by adding non-perturbative fields $\Phi^{(i)}$ with high charges that do not contribute to the leading order operators and setting the corresponding coefficient $\beta_i = 0$ in Eq.~\eqref{anomalyc} so the anomaly cancellation conditions remain satisfied.\\[2mm]
It is also worth noting that, since the charge patterns for the perturbative singlets are much more restrictive than those for the non-perturbative singlets, the space of entirely perturbative models is much smaller. For this reason,  it will be useful to search in Section~\ref{sec:res_pert} the space of models containing only perturbative singlets before tackling the full string FN models.

\section{Genetic algorithms and the string FN environment} \label{sec:ga}
As we have seen in the previous section, for many cases of interest the space of string FN models is too large to be amenable to systematic searches, so heuristic methods are required. Genetic algorithms (GAs) (see Refs.~\cite{goldberg2013genetic,Ruehle:2020jrk, katoch2021review} for reviews) have already proved to be extremely powerful in identifying physically viable models in a variety of contexts, including inflationary cosmology \cite{Abel:2022nje, Kamerkar:2022dfu}, BSM physics \cite{Allanach:2004my,Akrami:2009hp, Abel:2018ekz}, and string phenomenology~\cite{Abel:2014xta, Halverson:2019tkf,Abel:2021rrj, Abel:2021ddu, Cole:2021nnt, Abel:2023zwg}.
We start this section with a short review of GAs and then discuss the environment of string FN models.

\subsection{Genetic algorithms: a brief review}
Genetic Algorithms (GA) belong to a class of search algorithms motivated by natural evolution. Using mathematical parlance, the basic ingredients of a GA are summarised in the following diagram.
\begin{equation}\label{GA}
    [0,1]\stackrel{\pi_0}{\xleftarrow{\hspace{10mm}}} \mathbb{Z}_2^{N_{\rm bits}}\stackrel{\psi}{\xrightarrow{\hspace{10mm}}} E\stackrel{\nf}{\xrightarrow{\hspace{10mm}}}\mathbb{R}_{\leq0}
\end{equation}
Here, $E$ is the environment, also referred to as the space of phenotypes. Roughly speaking, for our applications, $E$ is the space of string FN models. On this space, we have a real-valued function $\nf$, the fitness function, which measures the `quality' of phenotypes. In our case, $\nf$ has a number of contributions (to be explained below) but the most important one measures the degree to which a certain string FN model reproduces the measured values of fermion masses and mixing.
Phenotypes are encoded as bit sequences, or genotypes, of length $N_{\rm bits}$ and the map $\psi$ is called the genotype-phenotype map. In general, this map does not need to be injective.\\[2mm] 
The genetic algorithm starts by selecting an initial population $P_0\subset\mathbb{Z}_2^{N_{\rm bits}}$, that is a genotype list of length $N_{\rm pop}$, randomly selected using a probability distribution $\pi_0$ (often, as it is also the case here, taken to be the flat distribution). The initial population is evolved in $N_{\rm gen}$ steps,
\begin{equation}
    P_0\longrightarrow P_1\longrightarrow\cdots\cdots\longrightarrow P_{N_{\rm gen}}\; ,
\end{equation}
where each individual step, $P_i\rightarrow P_{i+1}$, involves carrying out three types of operations: (i)~selection, (ii) cross-over, (iii) mutation. We briefly explain these operations in turn.
\begin{enumerate}
 \item[(i)] For selection, a probability distribution $\pi_i:P_i\rightarrow [0,1]$ is introduced on the $i^{\rm th}$~population. This is done such that the probability $p_i(b)$ of a genotype $b\in P_i$ increases with their fitness $\nf(\psi(b))$. There are several concrete realisations of selection, but here we adopt selection by ranking. Genotypes $b\in P_i$ are first ranked by their fitness and the probability $\pi_i$ depends linearly on the rank $r(b)$, specifically
 \begin{equation}
\pi_i(b)~=~ \frac{2}{(1+\alpha)N_{\rm pop}} \left( 1+\frac{N_{\rm pop}-r(b)}{N_{\rm pop}-1}(\alpha-1)\right)~.
\end{equation}
Here, $\alpha$ is a parameter, typically chosen in the range $\alpha\in [2,5]$, which can be used to adjust the degree of bias towards fit individuals. Using the distribution $\pi_i$, $N_{\rm pop}/2$ pairs of genotypes from $P_i$ are selected.
\item[(ii)] For each pair $(b_1,b_2)$ selected in the previous step, a cross-over is performed. This amounts to randomly selecting a position along the bit string and swapping the `tails' of $b_1$ and $b_2$ to the right of this position, thereby obtained a new pair $(\tilde{b}_1,\tilde{b}_2)$. Carrying this out for all pairs leads to a new population, $\tilde{P}_{i+1}$ of the same size $N_{\rm pop}$.
\item[(iii)] Finally, a random fraction, $p_{\rm mut}$, of bits in the population $\tilde{P}_{i+1}$ is flipped and this leads to $P_{i+1}$.
\end{enumerate}
When the evolution has been completed by iterating these steps, the `viable' states $b$ from all populations $P_0,\ldots ,P_{N_{\rm gen}}$ are collected. These are states with a sufficiently large fitness, $\nf(b)\geq \nf_{\rm term}$, large enough to guarantee they satisfy all desired properties. For our applications, the main property of viable string FN models is that they reproduce the measured fermion masses and mixing to sufficient accuracy. By generating many evolutionary cycles starting from different random initial populations $P_0$ and extracting viable states each time, it may be possible to obtain nearly complete lists of viable states, even for large environments~\cite{Abel:2021rrj,Abel:2021ddu,Abel:2023zwg}.

\subsection{The string FN environment} \label{sec:SFNenv}
In order to apply GAs to string FN models, we have to provide a concrete realisation of the various ingredients in the diagram~\eqref{GA}. We begin with the environment $E$.\\[2mm] 
First, we choose one of the partition vectors $\bn$ and charge patterns from Table~\ref{tab:10}. We then fix the numbers $N_\phi$ and $N_\Phi$ of perturbative and non-perturbative singlet fields. Given these choices, the associated string FN environment $E$ consists of all possible charges not already fixed in Table~\ref{tab:10} and all choices of singlet VEVs. Concretely, this includes the charges of the three $\bar{\bf 5}^I_{a_I,b_I}$ families (or, rather, the directions in which the $\bar{\bf 5}$ multiplets are charged), the charges of the perturbative singlets $\phi^{(\alpha)}_{c_\alpha,d_\alpha}$, where $\alpha=1,\ldots ,N_\phi$, and the charges $\bk^i$, where $i=1,\ldots ,N_\Phi$, of the non-perturbative singlets. To account for the VEVs in Eq.~\eqref{VEVs}, we also need to include the integers $p_\alpha$ and $\tilde p_i$. Altogether, the environment is then given by
\begin{equation}
    E = \{1,...,f\}^{(6+2N_\phi)}\times \{k_{\rm min},\ldots ,k_{\rm max}\}^{N_\Phi}\times \{1,\ldots ,p_{\rm max}\}^{N_\phi} \times \{\tilde p_{\rm min},\ldots , \tilde p_{\rm max}\}^{N_\Phi}
\end{equation}
In order to encode the elements of $E$ (states) as bit strings, we proceed in several steps. Firstly, consider the directions $(a_I,b_I)$ and $(c_\alpha,d_\alpha)$ in which the $\bar {\mathbf 5}$ multiplets and the $\phi$-singlets are, respectively, charged. These take values in the range $\{1,\ldots, f\}$ and we collate them into a single number in base $f$, then translate this into base $2$, obtaining a bit string of length $\ceil{\log_2(f^{6+2N_\phi})-1}$, where $\ceil{\cdot}$ is the ceiling function. \\[2mm]
For the $T$-charges, we choose $k_{\rm max}=k_{\min}+2^{N_k}-1$, for some positive integer $N_k$, such that each integer ${k^i}_a$ can be represented by $N_k$ bits. Finally, for the VEV integers, we choose $p_{\rm max}=2^{N_p}$ and $\tilde p_{\rm max}= \tilde p_{\rm min}+2^{N_{\tilde p}}-1$, such that each $p_\alpha$ and ${\tilde p}_i$ is represented by $N_p$ and $N_{\tilde p}$ bits, respectively. By concatenating these various pieces we obtain a bit string with length
\begin{equation} \label{eq:3.3}
  N_{\rm bits}=\ceil{\log_2(f^{6+2N_\phi})-1}+N_\Phi(f-1)N_k+N_\phi N_p+N_\Phi N_{\tilde{p}}\; .
\end{equation} 
The genotype-phenotype map $\psi:\mathbb{Z}_2^{N_{\rm bits}}\rightarrow E$ is defined by inverting this process. Due to the ceiling function in the first term in $N_{\rm bits}$, the part of the bit string that encodes the charges of the $\bar{\bf 5}$ families and perturbative singlets may sometimes produce, under inversion, a number that is too large; whenever this happens, we drop the first digit in the base-$f$ number after performing the inversion. \\[2mm]

\subsection{The fitness function}
Consider a bit string $b\in\mathbb{Z}_2^{N_{\rm bits}}$ and its associated phenotype $\psi(b)\in E$. In order to determine the physical viability of the model, we need to fix---at least approximately, in this first phase of GA search---the $\epsilon$-scale for the VEVs and the order-one coefficients. \\[2mm] % fix the physical model from this phenotype (and in preparation for the computation of the fitness function) we need to determine the reference scale $\epsilon$ and the order-one coefficients. First,  
The $\epsilon$-scale is set by first computing the $\epsilon$-order of magnitude for the eigenvalues of the up-type and down-type quark Yukawa matrices for generic values of the order-one coefficients. As in Section~\ref{sec:naturalness}, we define $\mathcal{M}^u = M^u (M^u)^\dagger$ and $\mathcal{M}^d = M^d (M^d)^\dagger$ where $M^u$, $M^d$ are the mass matrices for the up and, respectively, down quark sector. The eigenvalues of $\mathcal M^u$ and $\mathcal M^d$ satisfy  
\begin{equation}
\begin{aligned}
    \lambda_1^u &={\cal O}(\epsilon^{r_1^u-r_2^u-r_3^u})~,\qquad\lambda_2^u ={\cal O}(\epsilon^{r_2^u-r_3^u})~,\qquad\lambda_3^u ={\cal O}(\epsilon^{r_3^u})~,\\
        \lambda_1^d &={\cal O}(\epsilon^{r_1^d-r_2^d-r_3^d})~,\qquad\,\lambda_2^d ={\cal O}(\epsilon^{r_2^d-r_3^d})~,\qquad\,\lambda_3^d ={\cal O}(\epsilon^{r_3^d}),~
\end{aligned}
\end{equation}
where we have used the same conventions as in Section~\ref{sec:naturalness}.
We then match the their ratios to the observed quark mass ratios measured at the electroweak scale $M_Z$. Concretely, we define the four quantities $\epsilon_1$, $\epsilon_2$, $\epsilon_3$, $\epsilon_4$ by setting
\begin{equation}
    \frac{\lambda^{u}_1}{\lambda^{u}_2} \sim \epsilon_2^{r_1^{u}-2r_2^{u}} = \frac{m_u^2}{m_c^2}\; ,\qquad
    \; \frac{\lambda^{u}_2}{\lambda^{u}_3} \sim\epsilon_1^{r_2^{u}-2r_3^{u}} = \frac{m_c^2}{m_t^2}\;
\end{equation}
\begin{equation}
    \frac{\lambda^{d}_1}{\lambda^{d}_2} \sim\epsilon_4^{r_1^{d}-2r_2^{d}} = \frac{m_d^2}{m_s^2}\; ,\qquad
    \; \frac{\lambda^{d}_2}{\lambda^{d}_3} \sim \epsilon_3^{r_2^{d}-2r_3^{d}} = \frac{m_s^2}{m_b^2}\; .
\end{equation}

\vspace{2mm}
\noindent Generically, the values $\epsilon_i$ obtained from these four equations will be different. The $\epsilon$-scale is then chosen to minimise the logarithmic quadratic deviation, so it is given by the geometric mean of the $\epsilon_i$, that is, 
\begin{equation}\label{epsilon_optimised}
    \epsilon = \left( \epsilon_1\epsilon_2\epsilon_3\epsilon_4\right)^{1/4}~.
\end{equation} 

\noindent The next step is to choose the $18$ order-one coefficients for the up-type and down-type quark Yukawa matrices. Here we choose the order-one coefficients at $M_Z$-scale\footnote{As discussed in Section~\ref{sec:sugra_lang}, the coefficients in the Yukawa couplings should be of order-one at the GUT scale. However, during GA searches, we fix the coefficients to be of order one at the $M_Z$ scale so as to reduce the computational cost associated with the numerical integration involved in the renormalisation.}. We would also need to specify the two Higgs VEVs, $v^u$ and $v^d$. We start with a discussion of the latter, for which the following three relations are relevant: 
\begin{equation}
    m_t \approx v^u c_t \epsilon^{r_3^u}~,\qquad m_b \approx v^d c_b \epsilon^{r_3^d}~,\qquad v^2 = (v^u)^2 + (v^d)^2~,
\end{equation}
where $v$ is the measured Higgs VEV and $c_t$, $c_b$ are the order-one coefficients corresponding to the leading $\epsilon$-order entries along the diagonal in the two Yukawa matrices. To these equations, one can add the ratio $\tan \beta = v^u/v^d$ of the vacuum expectation values of the Higgs fields in the MSSM. Phenomenologically, $\tan \beta$ typically ranges from low (that is, $\tan\beta\gtrsim 1$) to moderate values, implying a large SUSY breaking scale, which aligns with the non-observation of superpartners at the LHC~\cite{CMS:2022goy, CMS:2024phk}. While $\tan\beta$ is qualitatively significant, it does not serve as a strict quantitative discriminator between models, so its value does not directly contribute to the GA fitness. However, the electroweak breaking scale, manifested by the Higgs VEV $v$, must be matched to account for observed phenomenological effects. We hence proceed as follows.\\[2mm]
The order-one coefficient $c_b$ for the bottom Yukawa coupling can be set to $1$ without loss of generality. This is so since the $\epsilon$-scale,  as determined above, is only accurate up to a factor of order one. The order-one coefficient~$c_b$ can then be absorbed into the definition of the $\epsilon$-scale.
The order-one coefficient $c_t$ for the top Yukawa coupling is then fixed by matching the Higgs VEV,
\begin{equation} \label{eq:3.1}
    v^2 = (v^u)^2 + (v^d)^2 \simeq \frac{m_t^2}{c_t^2\epsilon^{2r_3^{u}}} + \frac{m_b^2}{\epsilon^{2r_3^{d}}}\; .
\end{equation}
All other order-one coefficients are set to fixed random numbers (the same set of fixed random numbers for all models) in  the range $[0.5,3.0]$ and ensuing quark masses and mixings are compared with their values at the electroweak scale $M_Z$.\\[2mm]
Next, we should describe the fitness function $\nf :E\rightarrow \mathbb{R}_{\leq 0}$.  The fitness $\nf(b)$ of a state $b \in E$ is given by the sum of several contributions,
\begin{equation}\label{fitness_fn}
    \nf(b) = \nf_{\rm quark}(b) + \nf_{\rm texture}(b) + \nf_{\text{fine-tuning}}(b) %+ \nf_{\rm RG}(b) 
    + \nf_{\rm split\,type}(b) + \nf_{\rm anomaly}(b),
\end{equation}
described in detail in Appendix~\ref{sec:fitfunc}. Qualitatively, these contributions correspond to: 
\begin{itemize}
    \item $\nf_{\rm quark}(b)$ measures how well the model reproduces the measured quark masses and mixings.
    \item $\nf_{\rm texture}(b)$ checks whether the textures satisfy the naturalness conditions in Eq.~\eqref{eq:nat_cont}.
    \item $\nf_{\text{fine-tuning}}(b)$ measures how changes in the order-one coefficients will affect the measured quark masses and mixings. This is included so that the models found are not fine-tuned in the choice of order-one coefficients.
    \item $\nf_{\rm split\,type}(b)$ accounts for the additional rule that the charge choice $a=b$ for $\bar{\bf 5}_{a,b}$ is only allowed if the partition vector $\mathbf{n}$ satisfies $n_a>1$.
    \item $\nf_{\rm anomaly}(b)$ measures how well the  anomaly-cancellation conditions~\eqref{anomalyc} can be satisfied.
\end{itemize}
The $\nf_{\rm quark}(b)$ contribution is set such that two units of fitness correspond to an order of magnitude deviation away from any measured flavour parameter. A given model is deemed viable if the overall fitness is greater than $-20$, which roughly corresponds to allowing each of the $12$ flavour parameters in the quark sector to deviate, on average, by no more than one order of magnitude from their measured values. Models that pass this criterion typically retain their viability when the order-one coefficients are changed to any other random, generic choice.  
The contributions $\nf_{\rm split\,type}(b)$ and $\nf_{\rm anomaly}(b)$ are set such that any model that violates the above charge rule or the anomaly cancellation conditions is not accepted as viable.

\subsection{Implementation of the search strategy} \label{sec:strat}
The size of the environment grows exponentially with the number of FN singlets.
When the environment size is $\lesssim 10^8$ after eliminating the permutation redundancies, it is computationally feasible to perform a complete scan over the entire space. In those cases, we have separately performed a complete scan over all possible charge assignments and VEV choices. For the larger environments, we use the environment described above, implemented both in Mathematica and in C, coupled to the GA packages~\cite{GApackage} and \cite{GApackageC}, respectively. Since the charges of the perturbative singlets are much more constrained, we separately analyse the purely perturbative models and the models that also include non-perturbative insertions.
\vspace{2mm}
\begin{figure}[!tb]
    \centering
    \begin{minipage}{0.5\linewidth}
        \centering
        \includegraphics[width=\linewidth]{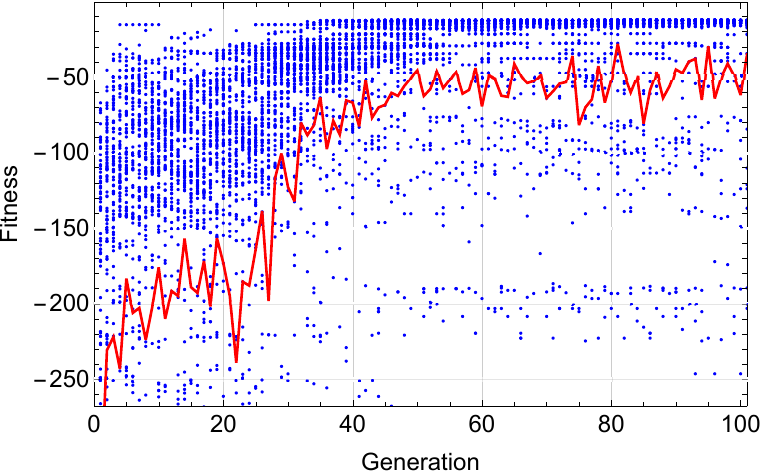}
    \end{minipage}
    \begin{minipage}{0.49\linewidth}
        \centering
        \includegraphics[width=\linewidth]{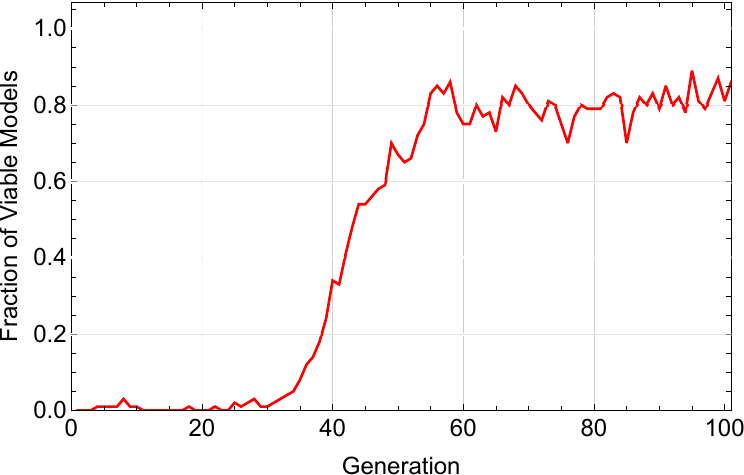}
    \end{minipage}
    \caption{The first plot shows the individual fitness (blue points) and the average population fitness (red curve) as a function of the generation, for a typical GA run in the string FN environment with  $\mathbf{n} = (1,1,3)$, $(N_\phi,N_\Phi) = (5,0)$ and the charges $\bm{10}_1$, $\bm{10}_2$, $\bm{10}_3$ and $\bm{\bar{5}}_{3,3}$ fixed. The second plot shows the fraction of viable models as a function of the generation.}
    \label{fig:ga_eg_run}  
\end{figure}

\begin{figure}[h]
\centering
\includegraphics[scale=0.6]{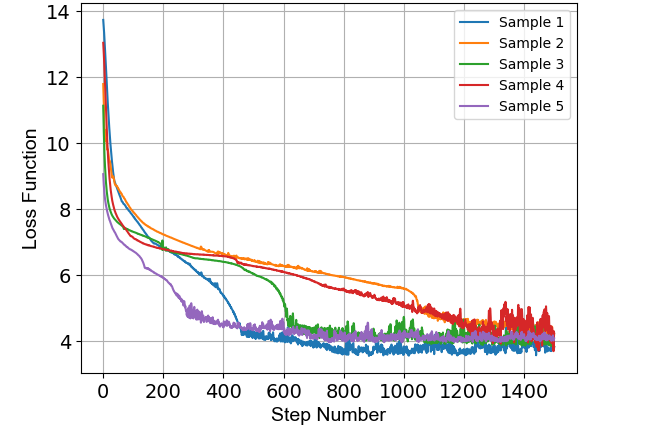}
\caption{Optimisation of the $\epsilon$-scale and order-one coefficients using the ADAM optimiser. The plot shows the evolution of the loss function during optimisation. The model corresponds to the partition vector $\mathbf{n} = (1,1,1,1,1)$ and the spectrum $\bm{10}_1$, $\bm{10}_2$, $\bm{10}_5$, $\bm{\bar{5}}_{1,2}$, $\bm{\bar{5}}_{1,2}$, $\bm{\bar{5}}_{1,2}$, $\bm{\bar{5}}^H_{4,5}$, $\bm{{5}}^H_{4,5}$, $\phi_{5,1}$, $\phi_{3,5}$, $\phi_{4,5}$, $\phi_{1,2}$, $\phi_{4,1}$. The five curves correspond to different random initialisations. The loss function~\eqref{eq:loss_func} is plotted on a log scale.
} \label{fig:optimise}
\end{figure}

\noindent Figure \ref{fig:ga_eg_run} illustrates a typical GA run (evolutionary cycle), displaying a clear increase in the population fitness, for a randomly generated initial population. 
The first phase of the search (whether comprehensive or heuristic) is intended to produce a list of charge patterns and Yukawa textures that approximately account for the values of the flavour parameters in the quark sector.\\[2mm]
In the second phase, we optimise the coefficients in the Yukawa couplings (assumed to be of order one at GUT scale) and the $\epsilon$-scale to obtain more accurate values for the Higgs VEV, the quark and charged lepton masses, and the CKM matrix.
An ADAM optimiser realised in JAX is used for this latter purpose. The RG-effects on the Yukawa couplings, described in Appendix~\ref{sec:rg}, are accounted for using the numerical integration module \texttt{diffrax}
to obtain the Yukawa couplings at the $M_Z$ scale.\\[2mm] 
More precisely, for the $\epsilon$-scale we search in the range $[0.05,0.9]$, while for the coefficients in the Yukawa couplings at GUT scale we search in the range $[0.5,3]$.
The loss function is set to be the $\ell^2$-norm of the relative deviation vector including the Higgs VEV, quark masses and mixings and charged lepton masses, where deviations are computed with respect to the experimentally-measured Standard Model quantities:
\begin{equation} \label{eq:loss_func}
    L(\delta) = \sqrt{\sum_\gamma \left(\frac{\left| \rho_{\gamma} - \rho_{\gamma,0} \right|}{\rho_{\gamma,0}(1+0.01\delta)} \right)^2}
\end{equation}
where $\rho_\gamma \in \left\{ v, m^u, m^d, m^c, m^s, m^t, m^b, m^e, m^\mu, m^\tau, |(V_{\rm CKM})_{ij}| \right\}$ are the parameters computed for the model, while $\rho_{\gamma,0}$ are the corresponding experimental values.
We use the stochastic variable $\delta$ taken from a uniform distribution $\delta \in [-1,1]$ to encourage exploration beyond local optima.
Given a large number of initial points in the search space, say, of order $10^5$, it is generally possible to optimise the $\epsilon$-scale and the set of order-one coefficients to obtain consistent masses and mixings. Figure~\ref{fig:optimise} illustrates the evolution of the loss function during optimisation for five random initial points in a particular string FN model.

\section{Models containing only perturbative singlets} \label{sec:res_pert}
We begin by searching for models that include only perturbative singlets since, for a given number of singlets, the resulting environment size in this configuration is substantially reduced compared to cases where both perturbative and non-perturbative singlets are considered, thus reducing the computational challenge. This setting also covers all the models containing non-perturbative singlets whose effects can be neglected.\\[2mm]
The results of the scans and GA searches after reducing redundancies are shown in Table~\ref{tab:pert_red}. The naturalness conditions discussed in Section~\ref{sec:gf} cannot be satisfied for a single FN scalar so we start with $N_\phi=2$ and cover all cases up to and including $N_\phi=5$. We do not consider cases with $N_
\phi>5$ as the computational cost becomes too large.
Whenever the reduced size of the environment (as a consequence of the permutation symmetries listed Table~\ref{tab:10}) is smaller or equal to $10^8$, we perform a complete scan.\\[2mm]
As expected, the number of viable models increases with the value of $N_\phi$. In particular, we were able to identify only one viable model with $N_\phi = 2$, while for larger values of $N_\phi$, the number of models grows roughly by one to two orders of magnitude for each additional perturbative singlet in the spectrum, irrespective of the partition vector $\bn$ and the charge assignments for the $\bm{10}^I$ and $\bar{\bm{5}}^H$ multiplets.
\vspace{2mm}
\begin{table}[!t]
\begin{center}
\begin{tabular}{|c|c|c|c||c|c|c|c|}
\hline
\multirow{2}{*}{$\mathbf{n}$} & \multirow{2}{*}{Fixed charges} & \multirow{2}{*}{$N_\phi$}  & Env & States & Full &\multirow{2}{*}{ Models} & Inequiv. \\
& &  & Size & visited & ~scan~ && spectra\\
\hline
\hline
% n = (1,1,1,1,1)
\multirow{4}{*}{$(1,1,1,1,1)$} & \multirow{4}{*}{$\bm{10}_1, \bm{10}_2, \bm{10}_5, \bm{\bar{5}}^H_{4,5}$} & 2 & $10^{9} $ &   & Yes& 0& 0\\
 & & 3 &  $10^{11} $ & $ 10^8$  & & 0 & 0\\
 & & 4 &  $ 10^{13}$ & $ 10^{8}$  & &301&4\\
 & & 5 &  $ 10^{16}$ & $ 10^{9}$  & &29213&289\\
\hline
% n = (1,1,1,2)
\multirow{4}{*}{$(1,1,1,2)$} & \multirow{4}{*}{$\bm{10}_1, \bm{10}_2, \bm{10}_4, \bm{\bar{5}}^H_{4,4}$} & 2 &  $10^{8} $ & & Yes & 0&0\\
& & 3  &  $10^{10} $ &  & Yes &98&  1\\
& & 4  &  $10^{12} $ & $  10^{8}$ & &18825& 55\\
& & 5  &  $10^{14} $ & $ 10^{8}$ &  &320557& 449\\
\hdashline
% n = (1,1,1,2)
\multirow{4}{*}{$(1,1,1,2)$} & \multirow{4}{*}{$\bm{10}_2, \bm{10}_3, \bm{10}_4, \bm{\bar{5}}^H_{1,4}$} & 2  &  $10^{8}$ & & Yes & 0&0\\
 & & 3 &  $10^{10} $ &  & Yes & 56& 1\\
 & & 4  & $10^{12} $ & $ 10^{8}$ & &11538& 128\\
 & & 5 & $10^{14} $ & $ 10^{8}$ & & 259175 & 1128\\
\hdashline
% n = (1,1,1,2)
\multirow{4}{*}{$(1,1,1,2)$} & \multirow{4}{*}{$\bm{10}_1, \bm{10}_2, \bm{10}_3, \bm{\bar{5}}^H_{1,4}$} & 2 &  $10^{8} $ & & Yes & 0&0\\
 & & 3  &  $10^{10} $ &  & Yes &  70&3\\
 & & 4  & $10^{12} $ & $ 10^{8}$ &  &8110& 63\\
 & & 5  & $10^{14} $ & $ 10^{8}$ &  &204148& 500\\
\hdashline
% n = (1,1,1,2)
\multirow{4}{*}{$(1,1,1,2)$} & \multirow{4}{*}{$\bm{10}_1, \bm{10}_3, \bm{10}_4, \bm{\bar{5}}^H_{1,2}$} & 2  &  $10^{8} $ & & Yes & 0&0\\
& & 3 &  $10^{10}$ &  & Yes &  0 &0\\
& & 4 & $10^{12}$ & $ 10^{8}$ &  & 0&0\\
& & 5 & $10^{14}$ & $ 10^{8}$ & & 0&0\\
\hline
% n = (1,1,3)
\multirow{4}{*}{$(1,1,3)$} & \multirow{4}{*}{$\bm{10}_1, \bm{10}_2, \bm{10}_3, \bm{\bar{5}}^H_{3,3}$} & 2  & $10^{6} $ & & Yes & 8&1\\
& & 3 & $10^{8} $ &  & Yes & 1218& 18\\
& & 4 & $10^{10} $ & & Yes & 22734&81\\
& & 5 & $10^{12} $ & $ 10^{8}$ & & 154532&234\\
\hline
\multirow{4}{*}{$(1,2,2)$} & \multirow{4}{*}{$\bm{10}_1, \bm{10}_2, \bm{10}_3, \bm{\bar{5}}^H_{3,3}$} & 2  & $10^{6}$ &  & Yes & 0&0\\
& & 3 & $10^{8}$ & & Yes & 0&0\\
& & 4 & $10^{10}$ & & Yes & 0&0\\
& & 5 & $10^{12}$ & $ 10^{8}$ & & 0&0\\
\hline
\hline
\end{tabular}
\caption{Search results for fully perturbative string FN models. The table includes the partition vector $\mathbf{n}$, the fixed charges for the $\bm{10}$ and $\bar{\bm{5}}^H$ multiplets, the number of singlets $N_\phi$, the size of the GA environment, the number of visited states in the GA search followed by an indication of whether a complete scan is feasible and was performed and, finally, the number of viable models and the number of inequivalent charge patterns found.
} \label{tab:pert_red}
\end{center}
\end{table}

\begin{figure}[h]
\centering
\includegraphics[scale=0.4]{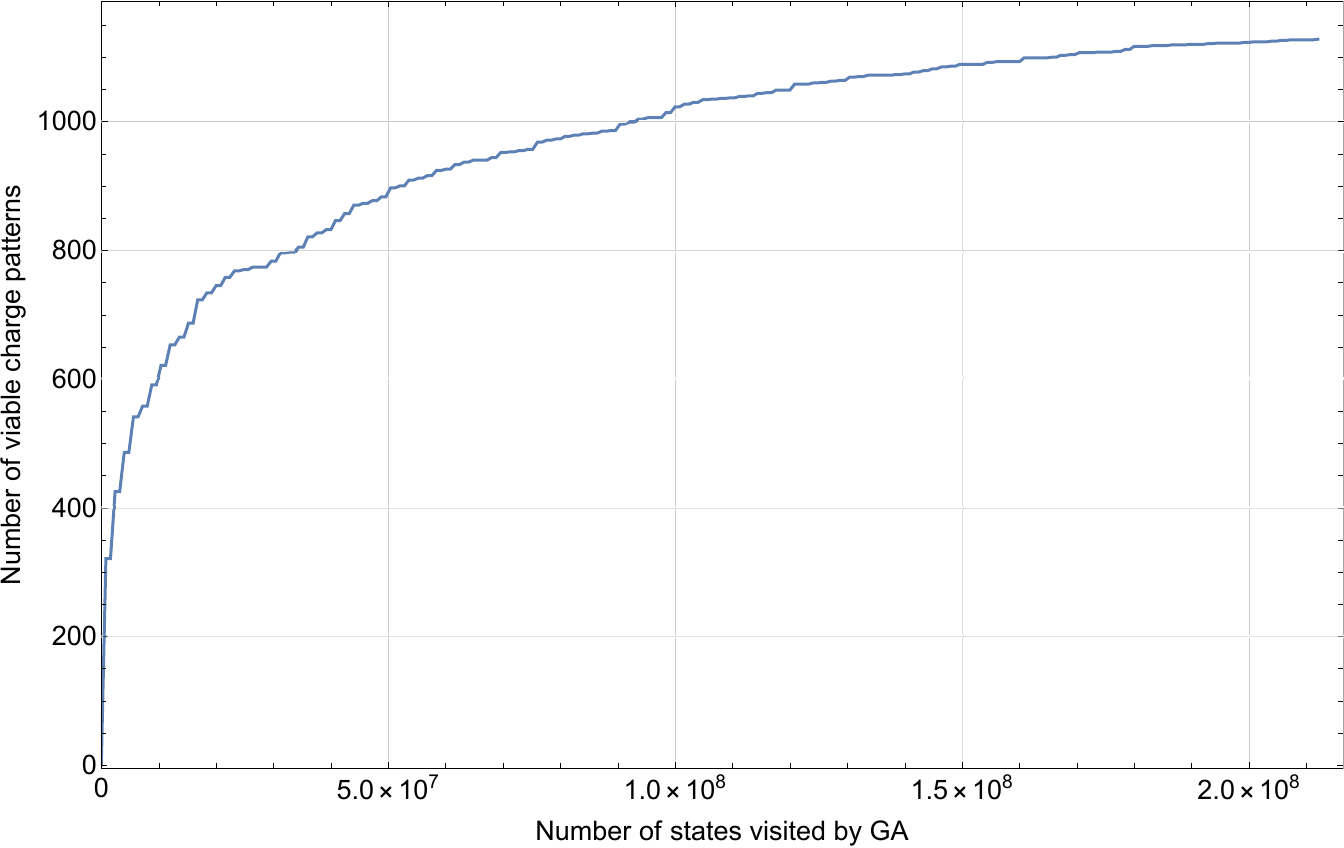}
\caption{Saturation plot for the number of viable (terminal) charge patterns as a function of the number of states visited by the GA. The models correspond to a partition vector $\mathbf{n} = (1,1,1,2)$ and fixed charges $\bm{10}_2$, $\bm{10}_3$, $\bm{10}_4$, $\bar{\bm{5}}_{1,4}^H$ with $N_\phi = 5$. The plot shows the total number of distinct charge patterns obtained in the GA search. } \label{fig:accum_graph}
\end{figure}

\noindent It is remarkable that the GA is capable of dealing with environments as large as $10^{15}$ states. To demonstrate the performance, we have included in Figure~\ref{fig:accum_graph} a saturation plot for the number of inequivalent (with respect to the permutation symmetries listed in Table~\ref{tab:10}) viable spectra (that is, charge patterns) identified as a function of the number of states visited. The plot suggests that almost all viable charge patterns in the given sector have been identified. From a high energy point of view, the set of viable charge patterns identified in the search may be translated into cohomological constraints for the compactification data. To this end, the viable charge patterns are more relevant than the full set of details describing viable models, which includes the VEV powers for the FN scalars. Below we exemplify the results by discussing in detail two viable models identified in the search.

\subsection{Example model with $\bn = (1,1,1,1,1)$}\label{sec:ex1_pert}
This model has already been mentioned in the Introduction, but we now present it in full detail. Its spectrum is given by
\begin{equation}
\bm{10}_1,~\bm{10}_2,~\bm{10}_5,~~\bm{\bar{5}}_{1,2},~\bm{\bar{5}}_{1,2},~\bm{\bar{5}}_{1,2},~~H^u_{4,5},~H^d_{4,5},~~\phi_{5,1},~\phi_{3,5},~\phi_{4,5},~\phi_{1,2},~\phi_{4,1}~
\end{equation}
and the $\epsilon$-orders of magnitude for the singlet VEVs are 
\begin{center}
\begin{tabular}{ccccc}
$\expval{\phi_{5,1}} \sim \epsilon^3$, & $\expval{\phi_{3,5}} \sim \epsilon^4$, & $\expval{\phi_{4,5}} \sim \epsilon$, & $\expval{\phi_{1,2}} \sim \epsilon^5$, & $\expval{\phi_{4,1}} \sim \epsilon^4$~.\\
\end{tabular}
\end{center}
This implies the leading-order singlet insertions
\begin{equation} \label{eq:yuk_eg_1_sing}
\begin{aligned}
\text{up sector:}&~~~~ \begin{pmatrix}
\!\!\!\phi_{5,1}\phi_{4,1}& \phi_{5,1}\phi_{1,2}\phi_{4,1} & ~\phi_{4,1}\\
\!\!\!~~\phi_{5,1}\phi_{1,2}\phi_{4,1}~ & ~\phi_{5,1}\phi_{1,2}^2\phi_{4,1}~ & ~\phi_{1,2}\phi_{4,1}\\
\!\!\!\phi_{4,1} & \phi_{1,2}\phi_{4,1} & ~\phi_{4,5}\\
\end{pmatrix}~,\\[4pt]
\text{down sector:}&~~~~ 
\begin{pmatrix}
\!\!\!\phi_{5,1}\phi_{3,5} & \phi_{5,1}\phi_{3,5} & ~\phi_{5,1}\phi_{3,5}\\
\!\!\!~\phi_{5,1}\phi_{3,5}\phi_{1,2}~ & ~\phi_{5,1}\phi_{3,5}\phi_{1,2}~ & ~\phi_{5,1}\phi_{3,5}\phi_{1,2}\\
\!\!\!\phi_{3,5} & \phi_{3,5} & ~\phi_{3,5}\\
\end{pmatrix}~,
\end{aligned}
\end{equation}
producing the Yukawa textures
\begin{equation} \label{eq:yuk_eg_1_text}
\text{up sector} \sim 
    \begin{pmatrix}
        \epsilon^7 & \epsilon^{12} & \epsilon^4\\
        \epsilon^{12} & \epsilon^{17} & \epsilon^9\\
        \epsilon^4 & \epsilon^9 & \epsilon\\
    \end{pmatrix},
    \quad
\text{down sector} \sim 
    \begin{pmatrix}
        \epsilon^7 & \epsilon^{7} & \epsilon^7\\
        \epsilon^{12} & \epsilon^{12} & \epsilon^{12}\\
        \epsilon^4 & \epsilon^4 & \epsilon^4\\
    \end{pmatrix}~.
\end{equation} 
Diagonalising these implies the following patterns for the up-quark, down-quark and charged lepton masses:
\begin{equation} \label{eq:yuk_eg_1_mass}
    (m_u, m_c, m_t)\sim (\epsilon^{17},\epsilon^7,\epsilon) ,\qquad (m_d, m_s,m_b),~(m_e,m_\mu,m_\tau)\sim (\epsilon^{12},\epsilon^7,\epsilon^4)~.
\end{equation}
For the GA fitness analysis, we computed the parameters $\epsilon = 0.44$, and $c_t = 3.0$. The resulting Higgs VEVs were given by $v^u = 130$ GeV, $v^d = 35.0$ GeV. With these choices, the total fitness value of the state was $-16.4$, hence the model passed as viable.\\[2mm] 
Finally, we carry out the optimisation of the $27$ order-one coefficients in both the quark and charged lepton sectors and of the $\epsilon$-scale. The optimised $\epsilon$-scale is $\epsilon = 0.554$ and the optimised order-one coefficients are:
\vspace{8pt}
\begin{equation}
\begin{array}{ccc}
~~\begin{pmatrix}
1.090 & 2.282 & 1.896\\
0.961 & 2.027 & 1.979\\
1.966 & 2.978 & 2.648\\
\end{pmatrix}~~ &~~ 
\begin{pmatrix}
1.379 & 1.843 & 0.947\\
2.708 & 1.726 & 2.063\\
1.530 & 2.526 & 0.680\\
\end{pmatrix}~~ &~~
\begin{pmatrix}
1.064 & 2.051 & 1.707\\
1.183 & 2.628 & 2.262\\
0.514 & 1.623 & 0.706\\
\end{pmatrix}~~\\[2pt]
\text{up-quark  sector} & \text{down-quark  sector} & \text{lepton sector}
\end{array}
\end{equation}
\vspace{4pt}

\noindent With these coefficients, the resulting values for the Higgs VEV, quark and lepton masses are listed in Table~\ref{table:ex1_pert}.

\begin{table}[!h]
\renewcommand{\arraystretch}{1.35}
\centering
\begin{tabular}{|c|c|c|c|}
\hline
\textbf{Higgs VEV} & \multicolumn{3}{c|}{\(\langle H \rangle\) = 174 GeV}  \\
\hline
\hline
\textbf{Quark} & \(m_u\) (MeV) & \(m_c\) (GeV) & \(m_t\) (GeV) \\
\hline
\textbf{Mass} & \(2.16\) & \(1.27\) & \(173\) \\
\hline
\textbf{Quark} & \(m_d\) (MeV) & \(m_s\) (MeV) & \(m_b\) (GeV) \\
\hline
\textbf{Mass} & \(4.70\) & \(93.9\) & \(4.18\) \\
\hline\hline
\textbf{Lepton} & \(m_e\) (MeV) & \(m_\mu\) (MeV) & \(m_\tau\) (GeV) \\
\hline
\textbf{Mass} & \(0.511\) & \(106\) & \(1.78\) \\
\hline
\end{tabular}
\caption{\sf Values for the Higgs VEV, quark and lepton masses predicted by the model in Section~\ref{sec:ex1_pert}.}
\label{table:ex1_pert}
\end{table}

\noindent The CKM matrix is given by
\begin{equation}
        \label{eqn:CKM_model_2}
        \left| V_{\text{CKM}} \right| \simeq \left(
            \begin{array}{ccc}
             0.970 & ~~0.242~~ & 0.00359 \\[2pt]
             0.242 & 0.969 & 0.0447 \\[2pt]
             0.00733 & 0.0443 & 0.999 \\
            \end{array}
            \right)\; .
\end{equation}
In this model, the relative deviations for the Higgs VEV, the quark and charged lepton masses are $<1 \%$. For the CKM matrix elements the relative deviations are slightly larger, going up to $\simeq 15\%$. This can be seen as a consequence of our two-step optimisation procedure, where the mass hierarchy is already addressed in the first step and further refined in the second.\\[2mm] 
It is likely that the model can be further optimised, however, in order to meaningfully do this one should also include a discussion of supersymmetry breaking into the RG analysis, something that goes beyond the scope of the present work. 

\subsection{Example model with $\bn = (1,1,3)$}\label{sec:ex2_pert}
We now look at a viable model with a different number of $U(1)$ symmetries (2 instead of 4), corresponding to the partition vector $\bn = (1,1,3)$. The model corresponds to the following spectrum:
\begin{equation}
\bm{10}_1,~\bm{10}_2,~\bm{10}_3,~~\bm{\bar{5}}_{1,2},~\bm{\bar{5}}_{3,3},~\bm{\bar{5}}_{1,2},~~H^u_{3,3},~H^d_{3,3},~~\phi_{3,1},~\phi_{2,3},~\phi_{1,1}~,
\end{equation}
with FN scalar VEVs of orders
\begin{equation*}
    \expval{\phi_{3,1}} \sim \epsilon^7, \quad \expval{\phi_{2,3}} \sim \epsilon^8, \quad \expval{\phi_{1,2}} \sim \epsilon^5~.
\end{equation*}
The leading-order singlet insertions are:
\begin{equation} \label{eq:yuk_eg_2}
\begin{aligned}
\text{up sector:}& \begin{pmatrix}
\phi_{3,1}^2 & \phi_{3,1}^2\phi_{1,2} & \phi_{3,1}\\
\phi_{3,1}^2\phi_{1,2} & \phi_{3,1}^2\phi_{1,2}^2 & \phi_{3,1}\phi_{1,2}\\
\phi_{3,1} & \phi_{3,1}\phi_{1,2} & 1\\
\end{pmatrix}~~ 
\text{down sector:}&\!\!\!\! 
\begin{pmatrix}
\phi_{3,1} & \phi_{2,3} & \phi_{3,1}\\
\phi_{3,1}\phi_{1,2}& \phi_{2,3}\phi_{1,2} & \phi_{3,1}\phi_{1,2}\\
1 & \phi_{2,3}^2\phi_{1,2} & 1\\
\end{pmatrix}
\end{aligned}
\end{equation}
giving the following Yukawa textures
\begin{equation} \label{eq:yuk_eg_2_text}
\text{up sector} \sim 
    \begin{pmatrix}
        \epsilon^{14} & \epsilon^{19} & \epsilon^7\\
        \epsilon^{19} & \epsilon^{24} & \epsilon^{12}\\
        \epsilon^{7} & \epsilon^{12} & 1\\
    \end{pmatrix},
    \quad
\text{down sector} \sim 
    \begin{pmatrix}
        \epsilon^7 & \epsilon^{8} & \epsilon^7\\
        \epsilon^{12} & \epsilon^{13} & \epsilon^{12}\\
        1 & \epsilon^{21} & 1\\
    \end{pmatrix}~,
\end{equation}
and the following textures of quark and charged lepton masses:
\begin{equation} \label{eq:yuk_eg_2_mass}
    (m_u, m_c, m_t) \sim (\epsilon^{24},\epsilon^{14},1) ,\qquad (m_d, m_s,m_b),~(m_e,m_\mu,m_\tau) \sim (\epsilon^{12},\epsilon^7,1)~.
\end{equation}
For the GA fitness analysis, we find the optimised values $\epsilon = 0.58$, $c_t = 0.99$ and Higgs VEVs $v^u = 174$ GeV, $v^d = 2.46$ GeV. The model has a GA fitness value of $-11.0$.\\[2mm]
In the final parameter optimisation phase, we find the optimal value $\epsilon = 0.692$ and the following optimised order-one coefficients:
\begin{equation}
\begin{array}{ccc}
~~\begin{pmatrix}
1.998 & 1.567 & 2.033\\
3.000 & 1.662 & 0.985\\
2.544 & 2.989 & 2.400\\
\end{pmatrix}~~ &~~ 
\begin{pmatrix}
2.340 & 0.501 & 1.292\\
2.362 & 0.500 & 0.925\\
2.576 & 3.000 & 0.509\\
\end{pmatrix}~~ &~~
\begin{pmatrix}
0.541 & 0.700 & 2.115\\
0.650 & 0.908 & 2.518\\
2.305 & 2.984 & 2.452\\
\end{pmatrix}~~\\[2pt]
\text{up-quark  sector} & \text{down-quark  sector} & \text{lepton sector}
\end{array}
\end{equation}

\noindent With these coefficients the flavour parameters resulting from this model are given in Table~\ref{table:ex2_pert}.
\begin{table}[!h]
\renewcommand{\arraystretch}{1.35}
\centering
\begin{tabular}{|c|c|c|c|}
\hline
\textbf{Higgs VEV} & \multicolumn{3}{c|}{\(\langle H \rangle\) = 174 GeV}  \\
\hline
%\textbf{Value} & \(174.14\) & & \\
\hline
\textbf{Quark} & \(m_u\) (MeV) & \(m_c\) (GeV) & \(m_t\) (GeV) \\
\hline
\textbf{Mass} & \(2.18\) & \(1.25\) & \(173\) \\
\hline
\textbf{Quark} & \(m_d\) (MeV) & \(m_s\) (MeV) & \(m_b\) (GeV) \\
\hline
\textbf{Mass} & \(4.68\) & \(94.0\) & \(4.18\) \\
\hline\hline
\textbf{Lepton} & \(m_e\) (MeV) & \(m_\mu\) (MeV) & \(m_\tau\) (GeV) \\
\hline
\textbf{Mass} & \(0.502\) & \(106\) & \(1.78\) \\
\hline
\end{tabular}
\caption{\sf Values for the Higgs VEV, quark and lepton masses predicted by the model in Section~\ref{sec:ex2_pert}.}
\label{table:ex2_pert}
\end{table}

\noindent The CKM matrix is given by
\begin{equation}
        \label{eqn:CKMNew}
        \left| V_{\text{CKM}} \right| \simeq \left(
            \begin{array}{ccc}
             0.970 & ~~0.242~~ & 0.00359 \\[2pt]
             0.242 & 0.969 & 0.0447 \\[2pt]
             0.00733 & 0.0442 & 0.999 \\
            \end{array}
            \right)\; .
\end{equation}

\noindent In this model, the relative deviations for the Higgs VEV, the quark and charged lepton masses are $<2\%$, while for the CKM matrix elements they are $<10\%$, except for the off-diagonal CKM element $|V_{31}|$ whose relative deviation is $\simeq 14.8 \%$. 

\section{Models with non-perturbative singlets} \label{sec:res_non-pert}
We now proceed to the discussion of models which include both perturbative and non-perturbative singlets. The charge patterns for the non-perturbative singlets are less constrained than their perturbative counterparts, being constrained only by Eq.~\eqref{c10}. For this reason, the environment size becomes significantly larger. Heuristic search methods such as GAs are, therefore, the only viable method of search.\\[2mm] 
Another consequence of the increased computational cost is that, in this setting, we will no longer aim to attain saturation in the number of viable models identified as a function of the number of GA runs. Instead, we will conduct approximately $10^8$ GA runs in each sector, as outlined in Table~\ref{tab:res_non-pert_5} and Table~\ref{tab:res_non-pert_122}. This number is comparable to the number of runs required to achieve saturation in the perturbative case. The resulting counts of inequivalent charge patterns are also similar, which, overall, represent only a tiny fraction of the total models explored.
\vspace{2mm}

\noindent To illustrate the results, we focus on two partition vectors $\bn$, namely the completely split case $\bn = (1,1,1,1,1)$ and the case $\bn = (1,2,2)$. The latter did not lead to any viable models in the fully perturbative setting but, as we will see, this situation changes with the inclusion of non-perturbative singlets. 

\subsection{Models with $\bn = (1,1,1,1,1)$}\label{sec:ex1_nonpert}
Following the arguments in Section~\ref{sec:gf}, in this setting naturalness conditions allow us to fix the following multiplet charges: $\bm{10}_1$, $\bm{10}_2$, $\bm{10}_5$, and $\bar{\bm{5}}_{4,5}$.
\vspace{2mm}
%
% Include summary of table
\begin{table}[ht]
\begin{center}
\begin{tabular}{|c|c|c||c|c|c|c|}
\hline
\multirow{2}{*}{$N_\phi$} & \multirow{2}{*}{$N_\Phi$} & Env. & States & Full & \multirow{2}{*}{Models} & Inequiv.\\
&& Size & Visited & Scan & & $q$-patterns\\
\hline
\hline
0 & 1 & $10^{9}$ &  & Yes & 0 & 0\\
0 & 2 & $ 10^{13}$ & $ 10^8$ & & 0 & 0\\
0 & 3 & $ 10^{18}$ & $ 10^8$ & & 0 & 0\\
0 & 4 & $ 10^{22}$ & $ 10^8$ & & 22 & 2\\
\hdashline
1 & 1 & $ 10^{11}$ & $ 10^8$ & & 0 & 0\\
1 & 2 & $ 10^{16}$ & $ 10^8$ & & 0 & 0\\
1 & 3 & $ 10^{20}$ & $ 10^8$ & & 648 & 107\\
1 & 4 & $ 10^{25}$ & $ 10^8$ & & 7377 & 2154\\
\hdashline
2 & 1 & $ 10^{14}$ & $ 10^8$ & & 0 & 0\\
2 & 2 & $ 10^{18}$ & $ 10^8$ & & 328 & 23\\
2 & 3 & $ 10^{23}$ & $ 10^8$ & & 6073 & 1490\\
2 & 4 & $ 10^{27}$ & $ 10^8$ & & 7354 & 1708\\
\hdashline
3 & 1 & $ 10^{16}$ & $ 10^8$ & & 76 & 5\\
3 & 2 & $ 10^{20}$ & $ 10^8$ & & 720 & 159\\
3 & 3 & $ 10^{25}$ & $ 10^8$ & & 892 & 149\\
\hdashline
4 & 1 & $ 10^{18}$ & $ 10^8$ & & 263 & 9\\
4 & 2 & $ 10^{23}$ & $ 10^8$ & & 2896 & 628\\
\hline
\end{tabular}
\caption{Results of the search for non-perturbative models with $\mathbf{n} = (1,1,1,1,1)$ and fixed multiplet charges $\bm{10}_1$, $\bm{10}_2$, $\bm{10}_5$, and $\bar{\bm{5}}_{4,5}$. The table includes the number of perturbative and non-perturbative fields $N_\phi$ and, respectively, $N_\Phi$, the size of the GA environment, the number of visited states in the GA search followed by an indication of whether a complete scan was performed, and finally, the number of viable models and the number of inequivalent charge patterns found, where equivalence is understood in the sense of the permutation symmetries listed in Table~\ref{tab:10}.} \label{tab:res_non-pert_5}
\end{center}
\end{table}

\noindent The results of the search are summarised in Table~\ref{tab:res_non-pert_5}. Except for the first case with $(N_\phi,N_\Phi) = (0,1)$ where an exhaustive scan is possible, the other environments are searched using GA with a fixed total number of visited states of $10^8$ states in each sector (which amounts to about 7000 core hours of computation). The results are presented in the same format as in the fully perturbative case.

\vspace{3mm}
\noindent To illustrate the results, it is useful to present an example. We do this for the sector $(N_\phi,N_\Phi) = (2,2)$, where we have found a model with the following spectrum
\begin{equation}
\bm{10}_1,~\bm{10}_2,~\bm{10}_5,~~\bm{\bar{5}}_{3,4},~\bm{\bar{5}}_{3,4},~\bm{\bar{5}}_{3,4},~~H^u_{4,5},~H^d_{4,5},~~\phi_{4,5},~\phi_{5,1},~\Phi^{(1)},~\Phi^{(2)}
\end{equation}
and charges for the two non-perturbative singlets $\Phi^{(1)}$ and $\Phi^{(2)}$ given by
\begin{equation}
    (k^i_a) = \begin{pmatrix}
    1 & -1 & 0 & -1 & 1\\
    1 & 1 & 0 & -1 & -1\\
    \end{pmatrix}
    \quad\begin{array}{l}\Phi^{(1)}\\ \Phi^{(2)}\end{array}
\end{equation}
The leading-order singlet insertions in the Yukawa matrices are then given by
\begin{equation} \label{eq:yuk_eg_3_sing}
\begin{aligned}
\text{up sector:}&~~~~ \begin{pmatrix}
\phi_{4,5}\phi_{5,1}^2~ & ~\phi_{4,5}^2\phi_{5,1}^2\Phi_1~ &~ \phi_{4,5}\phi_{5,1}\\
\phi_{4,5}^2\phi_{5,1}^2\Phi_1 ~&~ \phi_{4,5}^3\phi_{5,1}^2\Phi_1^2 ~&~ \phi_{4,5}^2\phi_{5,1}\Phi_1\\
\phi_{4,5}\phi_{5,1} ~&~ \phi_{4,5}^2\phi_{5,1}\Phi_1 ~&~ \phi_{4,5}\\
\end{pmatrix}~,\\[4pt]
\text{down sector:}&~~~~ 
\begin{pmatrix}
\phi_{5,1}\Phi_2 ~&~ \phi_{5,1}\Phi_2 ~&~ \phi_{5,1}\Phi_2\\
\phi_{4,5}\phi_{5,1}\Phi_1\Phi_2 ~&~ \phi_{4,5}\phi_{5,1}\Phi_1\Phi_2 ~&~ \phi_{4,5}\phi_{5,1}\Phi_1\Phi_2\\
\Phi_2 ~&~ \Phi_2 ~&~ \Phi_2\\
\end{pmatrix}~.
\end{aligned}
\end{equation}
Before proceeding further, let us first discuss how the anomaly cancellation condition in Eq.~\eqref{anomalyc} can be satisfied. A direct evaluation of $\mathbf{A}^{(5)}$ gives $\mathbf{A}^{(5)} = (3,3,3,3,3)\in\mathbb{Z}{\bf n}$, which is equivalent to the zero vector. A solution to the anomaly cancellation condition for this case is simply $(\beta_i) = (0,0)$, a result that imposes a strong constraint on the possible underlying string compactifications~\cite{Anderson:2012yf}.\\[2mm]
For the computation of Yukawa textures, we have identified the following $\epsilon$-powers for the singlet VEVs:
\begin{equation*}
    \expval{\phi_{4,5}} \sim \epsilon^3, \quad \expval{\phi_{5,1}} \sim \epsilon^8, \quad \expval{\Phi_1} \sim \epsilon^8, \quad
    \expval{\Phi_2} \sim \epsilon^7~.
\end{equation*}
We note that for this particular model, the VEVs of the perturbative singlet $\phi_{5,1}$ has to be stretched to a high power for the model to produce viable Yukawa textures. The resulting textures are given by:
\begin{equation} \label{eq:yuk_eg_3_text}
\text{up sector} \sim 
    \begin{pmatrix}
        \epsilon^{19} & \epsilon^{30} & \epsilon^{11}\\
        \epsilon^{30} & \epsilon^{40} & \epsilon^{22}\\
        \epsilon^{11} & \epsilon^{22} & \epsilon^3\\
    \end{pmatrix},
    \quad
\text{down sector} \sim 
    \begin{pmatrix}
        \epsilon^{15} & \epsilon^{15} & \epsilon^{15}\\
        \epsilon^{26} & \epsilon^{26} & \epsilon^{26}\\
        \epsilon^7 & \epsilon^7 & \epsilon^7\\
    \end{pmatrix}~.
\end{equation} 
which imply the following leading $\epsilon$-powers for the quark and charged lepton masses:
\begin{equation*}
    (m_u,m_c,m_t) \sim(\epsilon^{40},\epsilon^{19},\epsilon^3) ,\qquad (m_d,m_s,m_b),\, (m_e,m_\mu,m_\tau) \sim(\epsilon^{26},\epsilon^{15},\epsilon^7)
\end{equation*}
For the GA fitness analysis, we have computed the optimised parameters $\epsilon = 0.71$, $o_t = 2.85$ and the Higgs VEVs $v^u = 167.97$ GeV, $v^d = 13.48$ GeV. The fitness value was $-10.5$.\\[2mm]
In the second phase of optimisation we have identified an optimal $\epsilon$-scale at $\epsilon = 0.77$ and the following order-one coefficients:
\begin{equation}
\begin{array}{ccc}
~~\begin{pmatrix}
1.403 & 1.701 & 2.414\\
2.465 & 2.805 & 1.150\\
2.999 & 2.999 & 2.914\\
\end{pmatrix}~~ &~~ 
\begin{pmatrix}
1.389 & 1.737 & 1.554\\
0.564 & 0.500 & 2.995\\
2.997 & 2.962 & 0.500\\
\end{pmatrix}~~ &~~
\begin{pmatrix}
0.608 & 0.708 & 2.567\\
0.658 & 0.675 & 2.658\\
1.354 & 0.577 & 0.500\\
\end{pmatrix}~~\\[2pt]
\text{up-quark  sector} & \text{down-quark  sector} & \text{lepton sector}
\end{array}
\end{equation}
The resulting masses and mixing for this model are given in Table~\ref{table:ex1_non-pert}.
\begin{table}[!h]
\renewcommand{\arraystretch}{1.35}
\centering
\begin{tabular}{|c|c|c|c|}
\hline
\textbf{Higgs VEV} & \multicolumn{3}{c|}{\(\langle H \rangle\) = 175 GeV}  \\
\hline
\hline
\textbf{Quark} & \(m_u\) (MeV) & \(m_c\) (GeV) & \(m_t\) (GeV) \\
\hline
\textbf{Mass} & \(2.17\) & \(1.27\) & \(173\) \\
\hline
\textbf{Quark} & \(m_d\) (MeV) & \(m_s\) (MeV) & \(m_b\) (GeV) \\
\hline
\textbf{Mass} & \(4.75\) & \(93.8\) & \(4.18\) \\
\hline\hline
\textbf{Lepton} & \(m_e\) (MeV) & \(m_\mu\) (MeV) & \(m_\tau\) (GeV) \\
\hline
\textbf{Mass} & \(0.533\) & \(106\) & \(1.79\) \\
\hline
\end{tabular}
\caption{\sf Values for the Higgs VEV, quark and lepton masses predicted by the model in Section~\ref{sec:ex1_nonpert}.}
\label{table:ex1_non-pert}
\end{table}

\noindent The CKM matrix is given by
\begin{equation}
        \label{eqn:CKM_model_np_3}
        \left| V_{\text{CKM}} \right| \simeq \left(
            \begin{array}{ccc}
             0.970 & ~~0.242~~ & 0.00362 \\[2pt]
             0.242 & 0.969 & 0.0449 \\[2pt]
             0.00739 & 0.0445 & 0.999 \\
            \end{array}
            \right)\; .
\end{equation}
The flavour parameters obtained in this model are only a few percentages away from their measured values, the largest deviation corresponding to the off-diagonal CKM element $|V_{31}|$ of $\simeq 14.1\%$. Again, this is consistent with the mass hierarchy being more accurately accounted for in our two-step optimisation compared to the CKM matrix elements.

\subsection{Models with $\bn=(1,2,2)$}\label{sec:ex2_nonpert}
It is interesting to consider the partition vector $\mathbf{n} = (1,2,2)$ for which we have been unable to identify any viable perturbative models and investigate whether viable models with non-perturbative singlets can be found. To do this, we fix the multiplet charges $\bm{10}_1$, $\bm{10}_2$, $\bm{10}_3$, and $\bar{\bm{5}}^H_{3,3}$.
%
% Include summary of table
\begin{table}[h]
\begin{center}
\begin{tabular}{|c|c|c||c|c|c|c|}
\hline
$N_\phi$ & $N_\Phi$ & Env Size & States Visited & Full Scan & Models & $Q$-Patterns\\
\hline
\hline
0 & 1 & $ 10^{6}$ & & Yes & 0& 0\\
0 & 2 & $ 10^{8}$ & & Yes & 0& 0\\
0 & 3 & $ 10^{11}$ & $10^8$ & & 0& 0\\
\hdashline
1 & 1 & $ 10^{8}$ & & Yes & 0& 0\\
1 & 2 & $ 10^{10}$ & $ 10^8$ & & 0& 0\\
1 & 3 & $ 10^{13}$ & $ 10^8$ & & 0& 0*\\
\hdashline
2 & 1 & $ 10^{9}$ & $10^8$ & & 0& 0\\
2 & 2 & $ 10^{12}$ & $ 10^8$ & & 0& 0*\\
2 & 3 & $ 10^{15}$ & $ 10^8$ & & 0& 0*\\
\hdashline
3 & 1 & $ 10^{11}$ & $ 10^8$ & & 0& 0*\\
3 & 2 & $ 10^{14}$ & $ 10^8$ & & 0& 0*\\
3 & 3 & $ 10^{17}$ & $ 10^8$ & & 0& 0*\\
\hdashline
4 & 1 & $ 10^{13}$ & $ 10^8$ & & 25165 & 77\\
4 & 2 & $ 10^{16}$ & $ 10^8$ & & 317293 & 1338\\
\hline
\end{tabular}
\caption{Results of the search for non-perturbative models with $\mathbf{n} = (1,2,2)$ and fixed multiplet charges $\bm{10}_1$, $\bm{10}_2$, $\bm{10}_3$, and $\bar{\bm{5}}_{3,3}$. The table includes the number of perturbative and non-perturbative fields $N_\phi$ and, respectively, $N_\Phi$, the size of the GA environment, the number of visited states in the GA search followed by an indication of whether a complete scan was performed, and finally, the number of viable models and the number of inequivalent charge patterns found, where equivalence is understood in the sense of the permutation symmetries listed in Table~\ref{tab:10}.
} \label{tab:res_non-pert_122}
\end{center}
\end{table}

\noindent A summary of the search results in given in Table~\ref{tab:res_non-pert_122}. In particular, this setting differs from the completely-split $\bn = (1,1,1,1,1)$ setting in that no viable models could be found with $N_\phi \leq 3$, but this may be attributed to the rather limited search performed. In fact, looking at the viable models with $N_\phi=4$, many of these feature spectator singlets whose effects can be neglected at the level of analysis performed here. Such models belong, effectively, to the case $N_\phi \leq 3$.
\\[2mm]
To illustrate this discussion, we analyse a model from the sector $(N_\phi,N_\Phi) = (4,2)$ with the following spectrum:
\begin{equation}
\bm{10}_1,~\bm{10}_2,~\bm{10}_3,~~\bm{\bar{5}}_{1,3},~\bm{\bar{5}}_{1,3},~\bm{\bar{5}}_{1,3},~~H^u_{3,3},~H^d_{3,3},~~3\phi_{3,2},~\phi_{2,1},~\Phi^{(1)},~\Phi^{(2)}~.
\end{equation}
Although there are three singlets $\phi_{3,2}$ in the spectrum, it is clear that they can be treated as the same singlet. The charges of the non-perturbative fields $\Phi^{(1)}$ and $\Phi^{(2)}$ are
\begin{equation*}
    (k^i_a) = \begin{pmatrix}
    0 & ~2 & -2\\
    1 & -1 & ~~0\\
    \end{pmatrix}\quad
    \begin{array}{l}\Phi^{(1)}\\ \Phi^{(2)}\end{array}
\end{equation*}
With $\mathbf{A}^{(5)} = (6,3,6) \sim (3,-3,0)$, the anomaly-cancellation condition~\eqref{anomalyc} specialises to
\begin{equation}
    \begin{pmatrix}
        ~~0 & ~~1\\
        ~~2 & -1\\
        -2 & ~~0\\
    \end{pmatrix}
    \begin{pmatrix}
        \beta_1 \\ \beta_2
    \end{pmatrix} = 
    \begin{pmatrix}
        ~~3 \\ -3 \\ ~~0\\
    \end{pmatrix}\; .
\end{equation}
This equation is solved for $(\beta_1,\beta_2)= (0,3)$ and therefore the anomaly can be indeed cancelled by the Green-Schwarz mechanism. The leading-order singlet insertions are:
\begin{equation} \label{eq:yuk_eg_4}
\begin{aligned}
\text{up sector:}&~~~~ \begin{pmatrix}
\phi_{2,1}^2\phi_{3,2}^2 ~&~ \phi_{2,1}\phi_{3,2}^2 ~&~ \phi_{2,1}\phi_{3,2}\\
\phi_{2,1}\phi_{3,2}^2 ~&~ \phi_{3,2}^2 ~&~ \phi_{3,2}\\
\phi_{2,1}\phi_{3,2} ~&~ \phi_{3,2} ~&~ 1\\
\end{pmatrix}~,\\[4pt]
\text{down sector:}&~~~~ 
\begin{pmatrix}
\phi_{2,1}\phi_{3,2}\Phi_1 ~&~ \phi_{2,1}\phi_{3,2}\Phi_1 ~&~ \phi_{2,1}\phi_{3,2}\Phi_1\\
\phi_{3,2}\Phi_1 ~&~ \phi_{3,2}\Phi_1 ~&~ \phi_{3,2}\Phi_1\\
\Phi_1 ~&~ \Phi_1 ~&~ \Phi_1\\
\end{pmatrix}~.
\end{aligned}
\end{equation}
The leading contributions in the up-sector are entirely perturbative. However, this is not the case in the down-sector where the singlet $\Phi^{(1)}$ features, while $\Phi^{(2)}$ is a spectator field and its main purpose is to facilitate anomaly cancellation. The model fails to appear in the list of perturbative models from Section~\ref{sec:res_pert} despite the fact that $\Phi^{(1)}$ can be replaced by the product of $\phi_{2,3}^2$ of two perturbative singlets (assuming a spectrum that contains $\phi_{3,2}$, $\phi_{2,1}$,  and $\phi_{2,3}$ instead of the above spectrum). This happens because the leading singlet insertion in the first line of the down Yukawa matrix would, in this case, be  is $\phi_{2,1}\phi_{2,3}$ rather than  $\phi_{2,1}\phi_{3,2}\phi_{2,3}^2$. This discussion illustrates the relationship between the classes of purely perturbative models and the class of models with both types of singlets, as discussed in Section~\ref{sec:gf}. While it is  always possible to add non-perturbative singlets without significantly changing the phenomenology, it is not generally possible to replace every model with non-perturbative singlets by an equivalent purely perturbative one.
\vspace{3mm}

\noindent For the calculation of Yukawa textures, we have identified the following VEV patterns:
\begin{equation*}
    \expval{\phi_{3,2}} \sim \epsilon^7, \quad \expval{\phi_{2,1}} \sim \epsilon^6, \quad \expval{\Phi_1} \sim \epsilon^5
\end{equation*}
which lead to the Yukawa textures
\begin{equation} \label{eq:yuk_eg_4_text}
\text{up sector} \sim 
    \begin{pmatrix}
        \epsilon^{26} & \epsilon^{20} & \epsilon^{13}\\
        \epsilon^{20} & \epsilon^{14} & \epsilon^{7}\\
        \epsilon^{13} & \epsilon^{7} & 1\\
    \end{pmatrix},
    \quad
\text{down sector} \sim 
    \begin{pmatrix}
        \epsilon^{18} & \epsilon^{18} & \epsilon^{18}\\
        \epsilon^{12} & \epsilon^{12} & \epsilon^{12}\\
        \epsilon^5 & \epsilon^5 & \epsilon^5\\
    \end{pmatrix}~,
\end{equation} 
and the masses
\begin{equation*}
    (m_u,m_c,m_t) \sim(\epsilon^{26},\epsilon^{14},1) ,\qquad (m_d,m_s,m_b),\, (m_e,m_\mu,m_\tau) \sim(\epsilon^{18},\epsilon^{12},\epsilon^5)
\end{equation*}
For the GA fitness calculation we found the optimal parameters $\epsilon=0.62$, $c_t = 1.03$ and, furthermore, the Higgs VEVs  $v^u = 166$ GeV and $v^d = 122$ GeV. The fitness value of the model was $-5.1$.\\[2mm]
In the phase of optimisation, we find that $\epsilon = 0.40$ and the optimised order-one coefficients
\begin{equation}
\begin{array}{ccc}
~~\begin{pmatrix}
1.648 & 1.628 & 2.999\\
1.189 & 2.025 & 2.237\\
2.855 & 2.999 & 2.238\\
\end{pmatrix}~~ &~~ 
\begin{pmatrix}
2.698 & 2.254 & 1.096\\
1.255 & 1.248 & 0.501\\
3.000 & 2.699 & 0.500\\
\end{pmatrix}~~ &~~
\begin{pmatrix}
1.351 & 1.389 & 2.866\\
0.884 & 0.888 & 2.980\\
0.790 & 0.811 & 0.500\\
\end{pmatrix}~~\\[2pt]
\text{up-quark  sector} & \text{down-quark  sector} & \text{lepton sector}
\end{array}
\end{equation}
lead to the masses and Higgs VEV listed in Table~\ref{table:ex2_non-pert}.
\begin{table}[!h]
\renewcommand{\arraystretch}{1.35}
\centering
\begin{tabular}{|c|c|c|c|}
\hline
\textbf{Higgs VEV} & \multicolumn{3}{c|}{\(\langle H \rangle\) = 175 GeV}  \\
\hline
\hline
\textbf{Quark} & \(m_u\) (MeV) & \(m_c\) (GeV) & \(m_t\) (GeV) \\
\hline
\textbf{Mass} & \(2.20\) & \(1.27\) & \(173\) \\
\hline
\textbf{Quark} & \(m_d\) (MeV) & \(m_s\) (MeV) & \(m_b\) (GeV) \\
\hline
\textbf{Mass} & \(4.70\) & \(93.3\) & \(4.18\) \\
\hline\hline
\textbf{Lepton} & \(m_e\) (MeV) & \(m_\mu\) (MeV) & \(m_\tau\) (GeV) \\
\hline
\textbf{Mass} & \(0.521\) & \(107\) & \(1.84\) \\
\hline
\end{tabular}
\caption{\sf Values for the Higgs VEV, quark and lepton masses predicted by the model in Section~\ref{sec:ex2_nonpert}.}
\label{table:ex2_non-pert}
\end{table}

\noindent The CKM matrix is:
\begin{equation}
        \label{eqn:CKM_model_np_4}
        \left| V_{\text{CKM}} \right| \simeq \left(
            \begin{array}{ccc}
             0.970 & ~~0.241~~ & 0.00358 \\[2pt]
             0.241 & 0.969 & 0.0448 \\[2pt]
             0.00733 & 0.0443 & 0.999 \\
            \end{array}
            \right)\; .
\end{equation}

\noindent These flavour parameters are only a few percentages away from their measured values, the largest deviation being that of $|V_{31}|$ of $\simeq 14.7\%$. This model highlights the fact that some of the non-perturbative singlets may act as spectator fields (that is, they appear only in sub-leading contributions to the Yukawa couplings), however, their presence is needed to satisfy the anomaly cancellation conditions.

\section{Conclusion} \label{sec:con}
In this paper we have analysed a class of supersymmetric Froggatt-Nielsen (FN) models with multiple horizontal $U(1)$ symmetries and multiple SM singlets whose structure was motivated by heterotic string compactifications on smooth Calabi-Yau threefolds with split rank-five holomorphic vector bundles. Our approach mirrors the FN mechanism, however, key differences arise due to the string-theoretic origin which, in particular, implies a certain pattern of charges for the SM multiplets and the FN singlets, dividing the latter into two types (perturbative and non-perturbative singlets).
The first phase of exploration involved the identification of charge assignments for the SM multiplets and FN singlets, as well as singlet VEVs in terms of powers of a single scale~$\epsilon$, such that the observed mass hierarchies in the quark sector were reproduced qualitatively. We have employed a combination of comprehensive and heuristic (GA) search methods with a dedicated search for purely perturbative models, followed by a larger search for models involving both perturbative and non-perturbative FN singlets. The second phase involved an optimisation of the singlet VEVs and the order-one coefficients in the Yukawa matrices such that all flavour parameters reproduced their measured values within a few percentages. The optimisation was successful for the large majority of models generated in the first phase. \\[2mm]
In the fully perturbative sector we were able to identify several thousands of inequivalent spectra that can accommodate (with suitable choices of FN VEVs and order-one coefficients in the Yukawa matrices) the measured values for the Higgs VEV, the masses for the quark and charged leptons and the CKM matrix. While not exhaustive, we have run the GA until saturation, that is, up to the point where the number of new charge patterns becomes small in comparison with the number of charge patterns already identified. Remarkably, we were able to achieve saturation even for environments as large as $10^{16}$ while visiting only a $10^{-7}$ fraction of the available models.\\[2mm]
The environment in the non-perturbative sector was significantly larger, going up to a size of $10^{25}$ in certain sub-sectors and we have not attempted to reach any level of saturation during GA explorations. We have, however, identified several thousand of inequivalent charge patterns in this sector too. Some of the resulting models overlap with models from the perturbative sector, owing to the presence of spectator non-perturbative fields whose effects could be neglected. However, the examples discussed in Section~\ref{sec:res_non-pert} illustrate the fact that many of the models identified in this sector are qualitatively distinct from those in the perturbative sector. \\[2mm]
On the one hand, our results demonstrate that the class of string-inspired FN models is rich enough to accommodate all flavour physics. Although we have not studied the neutrino sector in this paper, the SM singlets present in these models can play the role of right handed neutrinos. We postpone the discussion of whether suitable masses can be generated in this sector to future work. 
On the other hand, compared to the size of the environments, the number of distinct charge patterns identified is very small, indicating that a careful engineering of $U(1)$ symmetries, multiplet charges and FN singlet VEVs is needed. Our bottom-up approach complements previous top-down analyses of heterotic Standard Models~\cite{Anderson:2011ns,Anderson:2012yf,Anderson:2013xka,Buchbinder:2013dna,Buchbinder:2014qda, Buchbinder:2014sya, Constantin:2018xkj}, including the recent computation of quark masses performed in Ref.~\cite{Constantin:2024yxh}. Our results indicate certain limits on the viable charge patterns which may be translated into cohomological constraints, such as the requirement that the three $\mathbf{10}$ multiplets have distinct $U(1)$ charges.
Furthermore, assuming the existence of cohomology formulae such as those studied in Refs.~\cite{Constantin:2018hvl, Klaewer:2018sfl, Brodie:2019dfx, Larfors:2019sie}, these observations may be translated into topological constraints on the allowed bundles. We also note that our investigation of non-perturbative models opens up an interesting range of phenomenological possibilities which include non-perturbative effects. These have, so far, not been explored systematically in a top-down string context due to the inherent difficulties associated with the computation of non-perturbative effects and, in particular, worldsheet instantons. \\[2mm]
Future work will attempt to bridge top-down and bottom-up approaches by identifying models from the databases in Ref.~\cite{Anderson:2013xka, Constantin:2018xkj} that can accommodate the observed flavour parameters. We also hope to integrate the neutrino sector into the analysis, potentially applying more advanced search methods such as Reinforcement Learning (RL), to navigate the more complex landscape of models.

\section*{Acknowledgements}
AC is supported by a Royal Society Dorothy Hodgkin Fellowship, grant~DHF/R1/231142. CSFT is supported by the Gould-Watson Scholarship. TRH is supported by the National Science Foundation under Cooperative Agreement PHY-2019786 (The NSF AI Institute for Artificial Intelligence and Fundamental Interactions, http://iaifi.org/). LTYL is supported by the Oxford-Croucher Scholarship. AL acknowledges support by the STFC consolidated grant ST/X000761/1. \\[2mm]
We would like to thank Jonathan Patterson for assistance with the Oxford Theoretical Physics computing cluster Hydra and Russell Jones for general technical support.

\newpage
\appendix   

\section{Naturalness conditions and charge-fixing constraints}\label{sec:nc_qfix}

In this Appendix, we outline the main arguments, relying on naturalness considerations, that lead to the charge-fixing constraints presented in Table~\ref{tab:10} of Section~\ref{sec:gf}. We summarize the key findings as follows:
\begin{itemize}
    \item Three distinct charges for the three $\bm{10}$ multiplets are necessary.
    \item If $\bar{\bf 5}^H_{a,b}$ is the multiplet containing the down-Higgs, the spectrum cannot contain both a ${\bf 10}_a$ and a ${\bf 10}_b$ multiplet. 
\end{itemize}

\subsection{Effects of order-one coefficients}
In our string-inspired FN models, additional order-one coefficients (controlled, in the presumed underlying string compactification, by geometric complex structure and K\"ahler moduli) are required to resolve the rank-deficiency degeneracies present in the FN mass textures. More explicitly, the rank-deficiency degeneracies arise when the structure of the FN-generated mass matrix has insufficient rank, leading to either zero masses or degenerate masses. To fix these issues, extra order-one coefficients (non-zero numbers) can be introduced into the mass matrix to break the degeneracies and lift the rank, ensuring that all fermions acquire distinct, non-zero masses, and a realistic mass hierarchy is generated.\\[2mm]
The order-one coefficients may introduce unwanted mass hierarchies via two mechanisms:
\begin{enumerate}
    \item The coefficients have hierarchically different magnitudes. Such effects may arise in certain asymptotic regions of the string moduli space. 
    \item Small differences in order-one coefficients can also generate a hierarchy. We refer to this as the fine-tuning case.
\end{enumerate}

\noindent In our analysis, we avoid the first mechanism by restricting the coefficients to be in the order-one range $[0.5, 3.0]$.\\[2mm] 
To illustrate the second mechanism, it is instructive to consider the matrix 
\begin{equation}
    B = 
    \begin{pmatrix}
        1 + \delta & 1 - \delta \\
        1 - \delta & 1 + \delta \\
    \end{pmatrix}
\end{equation}
with eigenvalues 
\begin{equation}
    (\lambda_1, \lambda_2) = (2, 2 \delta)
\end{equation}
If $\delta$ is a small parameter, a hierarchy between the two eigenvalues is generated solely through the small differences in the order-one coefficients. This second mechanism is particularly relevant for the up-type Yukawa matrices. Indeed, since the coupling $\bm{5}^H \bm{10}_I \bm{10}_J$ is symmetric in the family indices, the up-type Yukawa matrix (including singlet VEV insertions) will be symmetric; as such, small differences in the order-one coefficients may introduce unnatural hierarchies, as in the illustrative example presented above. We prove the following result.  

\begin{lem}
    To generate a natural FN structure, the three $\bm{10}$ families must have distinct $U(1)$ charges.
    \begin{proof}
        The argument is straightforward. If two $\bm{10}$ families have the same charge, the corresponding Yukawa submatrix would have four identical entries
        \begin{equation}
            \hat{\Lambda}_{[1,2]} \sim
            \begin{blockarray}{ccc}
                & \bm{10}^1_{a} & \bm{10}^2_{a} \\
                \begin{block}{c(cc)}
                    \bm{10}^1_{a} & * & * \\
                    \bm{10}^2_{a} & * & * \\
                \end{block}
            \end{blockarray}~.
        \end{equation}
        The only way to generate a hierarchy in such a case is by tuning the order-one coefficients to lift a zero eigenvalue, which we aim to avoid.
    \end{proof}
\end{lem}

\subsection{Off-diagonal entries in the up-Yukawa matrix}
In addition to the effects of order-one coefficients, we now discuss the effects of dominant off-diagonal entries in the up-Yukawa matrix.

\begin{lem} \label{lem:1.1}
    All off-diagonal terms in the up-type Yukawa matrix must be dub-dominant with respect to at least one diagonal entry in order to generate a natural hierarchy.
    \begin{proof}
        This follows from the naturalness conditions of Section~\ref{sec:naturalness}. Without loss of generality, we can focus on the up-type Yukawa matrix structure
        \begin{equation}
            \hat{\Lambda}^u \sim
            \begin{pmatrix}
                \epsilon^a~ & ~\epsilon^b~ & ~\epsilon^c \\
                \epsilon^b~ & ~\epsilon^d~ & ~1 \\
                \epsilon^c~ & ~1~ & ~\epsilon^e \\
            \end{pmatrix}
        \end{equation}
        where $a,b,c,d,e \in \mathbb{N}_{>0}$.\\[2mm] If $\lambda_1, \lambda_2, \lambda_3$ are the three eigenvalues of $\hat{\Lambda}^u$, it is straightforward to check that ${\rm tr}(\hat{\Lambda}^u)=\mathcal O(\epsilon^{{\rm min}(a,b,c)})$
        \begin{eqnarray}
\lambda_3+\lambda_2+\lambda_1&=&{\rm tr}(\hat{\Lambda}^u)={\cal O}(\epsilon^{{\rm min}(a,d,e)})\\
\lambda_3\lambda_2+\lambda_3\lambda_1+\lambda_2\lambda_1&=&\frac{1}{2}\left({\rm tr}(\hat{\Lambda}^u)^2-{\rm tr}((\hat{\Lambda}^u)^2)\right)={\cal O}(1)~,
\end{eqnarray}
        which is clearly a contradiction, since $a,d,e>0$. 
\end{proof}
\end{lem}

\begin{cor} \label{cor:1.1}
   If ${\bf 5}^H_{a,b}$ is the multiplet containing the up-Higgs, the spectrum cannot contain both the ${\bf 10}_a$ and ${\bf 10}_b$ multiplets. 
    \begin{proof}
        Suppose that the condition is not satisfied. Then the up-type Yukawa matrix contains the submatrix
        \begin{equation}
            \hat{\Lambda}^{(u)}_{a,b} \sim 
            \begin{pmatrix}
                *~ &~ 1 \\
                1~ & ~* \\
            \end{pmatrix}~.
        \end{equation}
        which is not allowed by Lemma~\ref{lem:1.1}.
    \end{proof}
\end{cor}

\section{Leading-order operators}\label{App:leading_ops}
In the first phase of exploration of the string FN landscape we identify viable Yukawa textures computed at the level of leading $\cG$-invariant operators, where $\cG\cong U(1)^{f-1}$.\\[2mm]
For models containing only perturbative singlets, we use the following algorithm to compute the leading-order operators. Recall that the $\cG$-charges of such singlets take the form
\begin{equation}
	\mathbf{q}(\phi_{a,b}) = {\mathbf e}_a- {\mathbf e}_b~,
\end{equation}
where $a=1,\ldots,f$. The singlet $\phi_{a,b}$ `moves' charges like a flow, taking one unit of charge from the $b$-th charge entry into the $a$-th entry. This perspective leads to to formulating the problem of finding leading-order operators as a Minimum Cost Maximum Flow (MCMF) problem. \\[2mm]
{\bfseries MCMF generalities.} The Minimum Cost Maximum Flow (MCMF) problem is a network optimization problem that seeks the least-cost distribution of a specified amount of flow in a network. The network is represented as a directed graph $G = (V, E)$, with:
\begin{enumerate}
    \item A set of source vertices $S \subset V$ and a set of sink vertices $T \subset V$.
\item A set of directed edges $(u, v) \in E$ between vertices $u, v \in V$, each edge having a capacity $c(u, v) > 0$ and a cost $t(u, v) > 0$.
\end{enumerate}

\noindent Given a demand $d_{s_i}$ for each source vertex $s_i \in S$ and a demand $d_{t_j}$ for each sink vertex $t_j \in T$, the objective is to determine a flow $f(u, v)$ that minimizes the total cost of flow across all edges:
\begin{equation}
    \sum_{(u, v) \in E} t(u, v) \cdot f(u, v)
\end{equation}
subject to the following constraints:
\begin{itemize}
    \item \textbf{Capacity constraint:} For each edge $(u, v) \in E$, the flow must not exceed the edge capacity: 
    \[
    f(u, v) \leq c(u, v).
    \]
    \item \textbf{Skew-symmetry constraint:} The flow must be skew-symmetric, meaning that for any pair of vertices $u, v \in V$:
    \[
    f(u, v) = -f(v, u).
    \]
    \item \textbf{Flow conservation constraint:} For each vertex $u \not\in S \cup T$, the total flow into the vertex must equal the total flow out, satisfying:
    \[
    \sum_{v \in V} f(u, v) = 0 =: d_u.
    \]
    For notational convenience, we write the net demand out of the vertex $u$ as $d_u$. For any vertex that is not a source or a sink, we must therefore have $d_u = 0$.
    \item \textbf{Source consistency constraint:} For each source vertex $s_i \in S$, the net flow out must match the source demand:
    \[
    \sum_{v \in V} f(s_i, v) = -d_{s_i}.
    \]
    \item \textbf{Sink consistency constraint:} For each sink vertex $t_j \in T$, the net flow into the sink must meet the sink demand:
    \[
    \sum_{v \in V} f(v, t_j) = d_{t_j}.
    \]
\end{itemize}

\noindent {\bfseries Implementation.} The problem of identifying leading-order operators can be translated into an MCMF problem as follows. First, we define the two charge vectors 
\begin{align}
{\mathbf q}^u= {\mathbf q}(H^u Q^Iu^J) &= {\mathbf q}(\bm{10}^I) + {\mathbf q}(\bm{10}^J) + {\mathbf q}(\bm{5}^H)\\
{\mathbf q}^d=	{\mathbf q}(H^d Q^Id^J) &= {\mathbf q}(\bm{10}^I) + {\mathbf q}(\bar{\bm{5}}^J) + {\mathbf q}(\bar{\bm{5}}^H)
\end{align}
and discuss them separately. Below we focus on ${\mathbf q}^u$, however, the other case is entirely analogous. \\[2mm]
Since $\cG$-charge vectors have $f$ entries, the corresponding set $V$ contains $f$ vertices, labelled as $1,\ldots,f$. The set $S$ consists of the positive entries of ${\mathbf q}^u$, while the set of negative entries forms the set of sinks $T$. Vertices $a$ and $b$ in $V$ are connected by the edge $b\rightarrow a$ only if the spectrum contains the perturbative singlet $\phi_{a,b}$. The demand vector is ${\mathbf d}=-{\mathbf q}^u$. Furthermore, the cost associated with the edge $b\rightarrow a$ is $t(b,a)=k$ if $\langle \phi_{a,b}\rangle = \epsilon^k$. The capacity $c(b,a)$ corresponds to the maximal number of allowed $\phi_{a,b}$-insertions. Finally, the aim is to determine the optimal flow function, with the understanding that $f(b,a)=n$ corresponds to having $n$-insertions of $\phi_{a,b}$ in the up Yukawa coupling. If the associated MCMF problem returns no solutions, this implies that there are no possible operators that match the SM operator charges up to the considered maximum number of insertions specified by edge capacity.

\begin{figure}
    \centering
    \resizebox{0.45\linewidth}{!}{
    \begin{tikzpicture}[auto, thick, scale=2, every node/.style={scale=1.5}]
    \node[circle, draw, fill=gray!30] (n1) at (90:2) {1};  
    \node[circle, draw, fill=gray!30] (n2) at (18:2) {2};   
    \node[circle, draw, fill=gray!30] (n3) at (-54:2) {3};  
    \node[circle, draw, fill=gray!30] (n4) at (-126:2) {4};
    \node[circle, draw, fill=gray!30] (n5) at (162:2) {5}; 

    % edges
    \draw[->, thick] (n1) -- (n4) node[midway, right] {$\phi_{4,1}$};
    \draw[->, thick] (n1) -- (n5) node[midway, above left] {$\phi_{5,1}$};
    \draw[->, thick] (n4) -- (n5) node[midway, left] {$\phi_{5,4}$};
    
    \end{tikzpicture}
    }
    \caption{Graph for the MCMF problem associated with finding the leading-order operator for the up-Yukawa coupling $\bm{10}_1\bm{10}_1\bm{5}^H_{4,5}$ with three perturbative singlets in the spectrum $\phi_{4,1}$, $\phi_{5,1}$, $\phi_{5,4}$ and ${\bn=(1,1,1,1,1)}$.} \label{fig:mcmf_graph}
\end{figure}
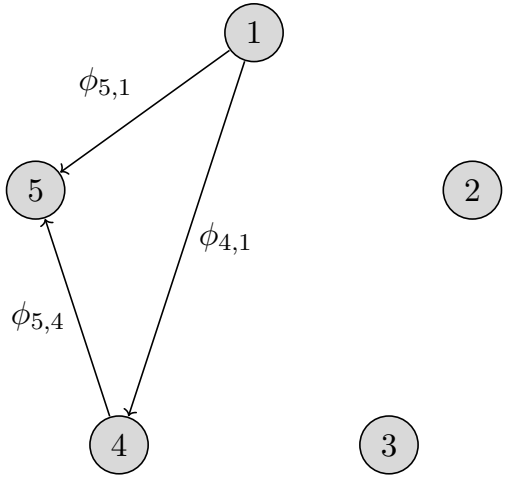

\vspace{2mm}
\noindent {\bfseries Example.} We now illustrate the method with model corresponding to $\bm{n} = (1,1,1,1,1)$. Suppose we have the up-Yukawa term $\bm{10}_1\bm{10}_1\bm{5}^H_{4,5}$ and the perturbative singlets $\phi_{4,1}$, $\phi_{5,1}$ and $\phi_{5,4}$ with
\begin{center}
\begin{tabular}{ccccc}
$\expval{\phi_{4,1}} \sim \epsilon^1$, & $\expval{\phi_{5,1}} \sim \epsilon^4$, & $\expval{\phi_{5,4}} \sim \epsilon^2$~.\\
\end{tabular}
\end{center}
The associated graph in the MCMF problem has five vertices, with edges $1\rightarrow 4$, $1 \rightarrow 5$ and $4\rightarrow 5$ having costs $t(1,4) = 1$, $t(1,5) = 4$ and $t(4,5) = 2$, respectively. The demand vector is ${\mathbf d} = -\mathbf{q}(\bm{10}_1\bm{10}_1\bm{5}^H_{4,5}) = (-2,0,0,1,1)$, indicating that vertex 1 is the sink with $d_1 = -2$ and vertices 4 and 5 are the sinks with $d_4 = d_5 = 1$. This is illustrated in Figure~\ref{fig:mcmf_graph}.

\noindent The solution to the associated MCMF problem is $f(1,4) = 2$, $f(1,5) = 0$ and $f(4,5) = 1$, such that
\begin{equation}
	{\mathbf q}(\phi_{4,1}^2\phi_{5,4}) = -{\mathbf q}(\bm{10}_1\bm{10}_1\bm{5}^H_{4,5})~.
\end{equation}

\noindent The algorithm for solving a MCMF problem is readily realised in Mathematica. We have also implemented the algorithm in C, achieving a speedup by a factor of $100$ in comparison with the approach of listing all $\cG$-invariant operators up to the given order and retaining the leading term. The algorithm complexity scales linearly with the number $N_\phi$ of singlets.
\vspace{3mm}

\noindent For models including non-perturbative singlets we adapt the above algorithm as follows. If $\{\Phi^{(i)}\}$ is the collection of non-perturbative singlets in the spectrum, we start by listing all possible combinations
\begin{equation}
    (\Phi^{(1)})^{i_1}(\Phi^{(2)})^{i_2}...(\Phi^{(N_\Phi)})^{i_{N_\Phi}}~,\qquad 0\leq i_1,\ldots i_{N_{\Phi}}\leq m~,
\end{equation}
for a fixed positive integer $m$. We then apply the above algorithm to each of these combinations of non-perturbative fields multiplied into the Yukawa coupling in question. The leading-order operator is then picked from the available solutions. This is computationally-viable provided that $m \leq 3$, which is what we have done in our searches.

\section{The fitness function} \label{sec:fitfunc}
The fitness function~\eqref{fitness_fn} of a state $b \in E$ in the GA environment consists of several contributions:
\begin{equation} \label{eq:fit_cont}
    \nf(b) = \nf_{\rm quark}(b) + \nf_{\rm texture}(b) + \nf_{\text{fine-tuning}}(b) + \nf_{\rm split\,type}(b) + \nf_{\rm anomaly}(b),
\end{equation}
We discuss each of these contributions in turn.\\

\noindent {\bfseries Masses and mixings.}
The main purpose of the fitness function is to measure how accurately the quark masses and mixings are reproduced. In the GA search, all Yukawa couplings are multiplied by arbitrary $\mathcal{O}(1)$ coefficients in order to break the rank-degeneracy of the Yukawa matrices. Since in this first phase the aim is to reproduce only qualitatively the observed hierarchies, we compute the logarithmic deviation of the computed quark masses and mixings from their corresponding experimental values at the $M_Z$ scale, setting:
\begin{equation}
\nf_{\rm quark} = - 2\sum_\gamma \min\left(\left|\log\left(\frac{\rho_\gamma}{\rho_{\gamma,0}}\right)\right|,10\right)~.
\end{equation}
Here, $\rho_\gamma \in \left\{ m^u, m^d, m^c, m^s, m^t, m^b, |(V_{\rm CKM})_{ij}| \right\}$ are the quark masses and mixings computed for the model, $\rho_{\gamma,0}$ are the corresponding experimental values and we have set a penalty bound of $-10$. The factor of 2 means that two units of fitness correspond to an order of magnitude deviation away from any measured flavour parameter.\\

\noindent {\bfseries Texture contributions.}
Following the discussion in Section~\ref{sec:gf}, the Yukawa textures  must be non-fine-tuned and the Higgs VEV must be fixed to its measured value at $M_Z$ scale. The texture contribution to the fitness function includes the following four terms:
\begin{equation}
    \nf_{\rm texture}(b) = -5\left[c_{\text{pow}}\nf_{\text{pow}} + c_{\epsilon}\nf_{\epsilon} + c_{O(1)}\nf_{O(1)} + c_{\text{range}}\nf_{\text{range}}\right]~,
\end{equation}
where the weights are set to 
$(c_{\text{pow}},c_{\epsilon},c_{O(1)},c_{\text{range}}) = (1,0.25,1,1)$. The overall factor of 5 is to make this contribution comparable with the other contributions in the fitness function in Eq.~\eqref{eq:fit_cont}. The various contributions are:
\begin{enumerate}
    \item[(i)] $\nf_{\text{pow}}$ accounts for the naturalness conditions described by Eq.~\eqref{eq:nat_cont}. $\nf_{\text{pow}}$ is zero when the naturalness conditions as satisfied, otherwise we take $\nf_{\text{pow}}$ to be the difference of the powers:
    \begin{equation}
        \nf_{\text{pow}} = \sum_{i=1,2}\left[\min\left(2r_{i+1}^{(u)} - r_i^{(u)},0\right) + \min\left(2r_{i+1}^{(d)} - r_i^{(d)},0\right) \right]
    \end{equation}
    \item[(ii)] $\nf_{\epsilon}$ is the geometric standard deviation of the $(\epsilon_1,\epsilon_2,\epsilon_3,\epsilon_4)$-distribution from which the optimised scale $\epsilon$ is obtained via Eq.~\eqref{epsilon_optimised},
        \begin{equation}
            \nf_{\epsilon} = \exp \left[ \frac{1}{4} \sum_{i=1}^4 \log\left(\frac{\epsilon_i^2}{\epsilon^2}\right)  \right]^{1/2}~.
        \end{equation}
    \item[(iii)] $\nf_{O(1)}$ measures whether the coefficient $c_t$ calculated from Eq.~\eqref{eq:3.1} is of order one and is real. If $c_t$ is real but outside of the range $[0.5,3]$, we set $\nf_{O(1)}$ to be the logarithm of the ratio between the bound and $c_t$ as follows:
    \begin{equation} \label{eq:o1}
        \nf_{O(1)} = \begin{cases}
            0 \quad &\text{if} \quad 0.5<c_t< 3\\
            \log\left(c_t/3\right)\quad &\text{if} \quad 3<c_t \\
            \log\left(0.5/c_t\right)\quad &\text{if} \quad c_t < 0.5\\
        \end{cases}~.
    \end{equation}
    If $c_t$ imaginary, the correct Higgs VEV cannot be obtained, in which case $\nf_{O(1)} = -c_t^2$.
    \item[(iv)] $\nf_{\text{range}}$ accounts for the ranges of the singlet VEVs. We typically expect the singlets to lie between the following ranges: $\expval{\phi} \in [0.005,0.5]$ and $\expval{\Phi} = [0.0005,0.05]$. However, since we calculate the $\epsilon$-scale after specifying the VEV powers, the actual VEV of the singlets may be outside the range as specified. We therefore penalise models that give a singlet VEV outside the range by the following fitness contribution,
    \begin{equation}
        \nf_{O(1)} = \sum_\alpha \nf_{O(1),\phi^{(\alpha)}} + \sum_i \nf_{O(1),\Phi^{(i)}}
    \end{equation}
    where the individual contributions are,
    \begin{equation}
        \nf_{O(1),\phi^{(\alpha)}} = \begin{cases}
            0 \quad &\text{if} \quad 0.005< \expval{\phi^{(\alpha)}} < 0.5\\
            \log\left(\expval{\phi^{(\alpha)}}/0.5\right)\quad &\text{if} \quad 0.5<\expval{\phi^{(\alpha)}} \\
            \log\left(0.005/\expval{\phi^{(\alpha)}}\right)\quad &\text{if} \quad \expval{\phi^{(\alpha)}} < 0.005\\
        \end{cases}~,
    \end{equation}
    and
    \begin{equation}
        \nf_{O(1),\Phi^{(i)}} = \begin{cases}
            0 \quad &\text{if} \quad 0.0005< \expval{\Phi^{(i)}} < 0.05\\
            \log\left(\expval{\Phi^{(i)}}/0.05\right)\quad &\text{if} \quad 0.05<\expval{\Phi^{(i)}} \\
            \log\left(0.0005/\expval{\Phi^{(i)}}\right)\quad &\text{if} \quad \expval{\Phi^{(i)}} < 0.0005\\
        \end{cases}~.
    \end{equation}
\end{enumerate}

% The O(1) coefficient contributions
\noindent {\bfseries Order-one tuning contributions.}
The order-one coefficient fine-tuning contributions to quark mixing are not considered in the fine-tuning analysis of Yukawa textures. This however should be accounted for, as by the fitness contribution $\nf_{\text{fine-tuning}}$. Following Ref.~\cite{Ross:2017kjc}, we have taken the fine-tuning fitness contribution $f_{\text{fine-tuning}}$ to be the logarithmic numerical differentiation of the SM quantities with respect to the order-one coefficients:
\begin{equation}
\nf_{\text{fine-tuning}} = -\frac{1}{10} \sum_{i} \text{max}_j\left|\frac{\delta \log\rho_i}{\delta \log o_j}\right|
\end{equation}
where $\rho_i \in \left\{\expval{H},m_i,|V_{ij}|\right\}$ are the set of Standard Model quantities. The overall factor of $1/10$ is to allow this contribution to be comparable to the other contributions in Eq.~\eqref{eq:fit_cont}.
\vspace{2mm}

% Non-abelian contributions
\noindent{\bfseries Contributions from the charge constraint.}
Whenever the spectrum contains charge patterns that are forbidden, such as having $e_a = e_b$ for the multiplets $\bm{\bar{5}}$ in cases where the partition vector ${\mathbf n}=(1,1,1,1,1)$, the contribution $\nf_{\rm split\,type}$ is evaluated as:
\begin{equation}
    \nf_{\rm split\, type} = -20 \times N_{\text{violations}}
\end{equation}
where $N_{\text{violations}}$ is the number of forbidden charge patterns in the spectrum. This contribution choice is shown to have better convergence effects in the GA search than simply setting $\nf_{\rm split\,type} = -20$ whenever such charge constraints are violated.
\vspace{2mm}

% GS anomaly contributions
\noindent{\bfseries Anomaly cancellation contributions.}
The term $\nf_{\rm anomaly}$ accounts for the failure to satisfy the anomaly cancellation condition in Eq.~\eqref{anomalyc}. To eliminate all models which fail to satisfy the anomalous cancellation condition from the viable model list, $\nf_{\rm anomaly} = 0$ if an integer solution of $\beta_i$ can be found and $\nf_{\rm anomaly} \leq -20$ for models with no integer solutions of $\beta_i$. Similar to the charge constraint contribution, we introduce continuous-varying contribution for the cases with no integer solutions of $\beta_i$ to achieve better convergence in the GA search. In particular, when $\beta_i$ has no integer solutions but a solution $\beta_i \in \mathbb{R}^{N_T}$ could be found, we set
\begin{equation}
    \nf_{\rm anomaly} = -20 \times \left[\min\left(\sum_i \left[\text{round}(\beta_i) - \beta_i\right]-1, 0 \right) + 1 \right]
\end{equation}
where round stands for rounding to the nearest integer. If no solutions can be found for Eq.~\eqref{anomalyc}, then
\begin{equation}
    \nf_{\rm anomaly} = 
    -20 \times \left[\min_{\beta_i}\left(||\mathcal{A}_a - \beta_i k^i_a||-1, 0 \right) + 1 \right]
\end{equation}
where the minimal $\beta_i \in \mathbb{R}^{n_T}$ is the approximate solution to the original linear system found by the normal equation (using the notation $k\indices{^i_a}$ to indicate the matrix elements),
\begin{equation}
    \beta_i = (k^Tk)\indices{^j_i}(k^T)\indices{^a_j} \mathcal{A}_a
\end{equation}
in the method of ordinary least squares.

\section{Crude renormalisation scheme} \label{sec:rg}
%In this section we outline the steps taken in crude renormalisation scheme. 
The MSSM one-loop RG equations~\cite{Castano:1993ri} are given by:
\begin{equation}
\frac{d\Lambda^{u,d,e}}{dt} = \frac{1}{16\pi^2} \beta^{u,d,e}~,\qquad
\frac{dg_a}{dt} = \frac{g_a^3}{16\pi^2}\beta_a
\end{equation}
where $t = \log(M/M_0)$, $M_0$ is the reference scale, $(\beta_a) = (33/5,1,-3)$ for $U(1)$, $SU(2)$ and $SU(3)$, respectively, and
\begin{align}
\beta^u &= \left( 3\tr(\Lambda^u\Lambda^{u\dagger}) + 3\Lambda^u\Lambda^{u\dagger} + \Lambda^d\Lambda^{d\dagger} -\frac{16}{3} g_3^2 -3g_2^2 -\frac{13}{15}g_1^2 \right) \Lambda^u\\
\beta^d &= \left( \tr(3\Lambda^d\Lambda^{d\dagger} + \Lambda^e\Lambda^{e\dagger}) + 3\Lambda^d\Lambda^{d\dagger} + \Lambda_u\Lambda_u^\dagger -\frac{16}{3} g_3^2 -3g_2^2 -\frac{7}{15}g_1^2 \right) \Lambda^d\\
\beta^e &= \left( \tr(3\Lambda^d\Lambda^{d\dagger} + \Lambda^e\Lambda^{e\dagger}) + 3\Lambda^e\Lambda^{e\dagger}  -3g_2^2 -\frac{9}{5}g_1^2 \right) \Lambda^e
\end{align}
We take the following radical approach. We neglect the effects of $g_1$ and $g_2$ (only retaining strong effects) and the effects of the $(3,3)$-entries of the Yukawa matrices, i.e.\
\begin{equation}
\Lambda^u \simeq \begin{pmatrix}
0~ & 0~ & 0\\
0~ & 0~ & 0\\
0~ & 0~ & y_t\\
\end{pmatrix}
, \quad
\Lambda^d \simeq \begin{pmatrix}
0~ & 0~ & 0\\
0~ & 0~ & 0\\
0~ & 0~ & y_b\\
\end{pmatrix}
, \quad
\Lambda^e \simeq \begin{pmatrix}
0~ & 0~ & 0\\
0~ & 0~ & 0\\
0~ & 0~ & y_\tau\\
\end{pmatrix}~.
\end{equation}
We insert these in the above beta functions. Integrating the $g_3$ one-loop RG equation and redefining the variable $t \mapsto t/16\pi^2$ gives
\begin{equation}
g_3^2 = \frac{g_{3,0}^2}{1-2g_{3,0}^2\beta_3t}~,
\end{equation}
where $t=0$ corresponds to $M = M_Z$, $t\simeq 0.201$ corresponds to $M = M_{\text{GUT}}$, $\beta_3 = -3$, $g_3^2 = 4\pi\alpha_3$ and $\alpha(M_Z) \simeq 0.1155$. The differential equations for the Yukawa matrices then become:
\begin{align} \label{eq:rg.1}
\frac{dy_t}{dt} &{=}  ( 6y_t^2 {+} y_b^2 {-} \frac{16}{3}g_3^2 ) y_t, \; &\frac{dy_b}{dt} &{=}  ( 6y_b^2 {+} y_t^2 {+} y_\tau^2 {-} \frac{16}{3}g_3^2 ) y_b \; &\frac{dy_\tau}{dt} &{=}  ( 3y_b^2 {+} 4y_\tau^2 ) y_\tau\\
\frac{d\Lambda^u_{3,i}}{dt} &{=}  ( 6y_t^2 {+} y_b^2 {-} \frac{16}{3}g_3^2 ) \Lambda^u_{3,i}, \; &\frac{d\Lambda^d_{3,i}}{dt} &{=}  ( 6y_b^2 {+} y_t^2 {+} y_\tau^2 {-} \frac{16}{3}g_3^2 ) \Lambda^d_{3,i}, \; &\frac{d\Lambda^e_{3,i}}{dt} &{=}  ( 3y_b^2 {+} 4y_\tau^2 ) \Lambda^e_{3,i}\\
\frac{d\Lambda^u_{i,j}}{dt} &{=}  ( 3y_t^2 {-} \frac{16}{3}g_3^2 ) \Lambda^u_{i,j}, \; &\frac{d\Lambda^d_{i,j}}{dt} &{=}  ( 3y_b^2 {+} y_\tau^2 {-} \frac{16}{3}g_3^2 ) \Lambda^d_{i,j}, \;
&\frac{d\Lambda^e_{i,j}}{dt} &{=}  y^2_\tau \Lambda^e_{i,j}
\end{align}
where we have denoted $\Lambda^u_{3,3} = y_t$, $\Lambda^d_{3,3} = y_b$ and $\Lambda^e_{3,3}$. The coupled differential equations in Eq.~\eqref{eq:rg.1} have no known closed solutions (a recursive solution exists in \cite{Auberson:1999kv,MSSMWorkingGroup:1998fiq}) but can be solved numerically. The other Yukawa couplings $x=u,d,e$ can be solved using:
\begin{equation}
\frac{\Lambda^x_{m,n}}{(\Lambda_0)^x_{m,n}} = \exp\left[ \int_0^{t} dt\left( k_{t}y_t^2 + k_{b}y_b^2 + k_\tau y_\tau^2 - \frac{16}{3}g_3^2 )\right) \right]
\end{equation}
where $i,j \in \{1,2\}$, $x$ is the family index $x = u,d,e$ for the up-, down- and electron-Yukawa couplings, respectively, and
\begin{equation}
    (k_t,k_b,k_\tau) = \begin{cases}
        (6,1,0) \quad &\text{if} \quad x=u,~(n=3\text{ and }m=i)\text{ or }(n=i\text{ and }m=3)\\
        (3,0,0) \quad &\text{if} \quad x=u,~(m=i\text{ and }n=j)\\
        (1,6,1) \quad &\text{if} \quad x=d,~(n=3\text{ and }m=i)\text{ or }(n=i\text{ and }m=3)\\
        (0,3,1) \quad &\text{if} \quad x=d,~(m=i\text{ and }n=j)\\
        (0,3,4) \quad &\text{if} \quad x=e,~(n=3\text{ and }m=i)\text{ or }(n=i\text{ and }m=3)\\
        (0,0,1) \quad &\text{if} \quad x=e,~(m=i\text{ and }n=j)\\
    \end{cases}~.
\end{equation}

\bibliography{flavour_hetsm_qsec}
\bibliographystyle{inspire}

\end{document}